\documentclass{emulateapj}
\usepackage{amsmath}
\usepackage{natbib}
\usepackage{txfonts}
\usepackage{graphicx,color}

\begin{document}
\shorttitle{Three-dimensional horseshoe drag}
\shortauthors{F. S. Masset \& P. Ben\'\i tez-Llambay}
\title{Horseshoe drag in three-dimensional globally isothermal disks}
\author{F.\ S.\ Masset\altaffilmark{1}} 
\affil{Instituto de Ciencias F\'\i sicas, Universidad
Nacional Aut\'onoma de M\'exico (UNAM), Apdo. Postal 48-3,
62251-Cuernavaca, Morelos, M\'exico}
\email{masset@icf.unam.mx}
\altaffiltext{1}{Send offprint requests to masset@icf.unam.mx}
\author{P.  Ben\'\i tez-Llambay\altaffilmark{2}}
\affil{Instituto de Astronom\'\i a Te\'orica y Experimental, IATE (CONICET),
   Observatorio Astron\'omico, Universidad Nacional de C\'ordoba, 
   Laprida 854, X5000BGR, C\'ordoba, Argentina}
\email{pbllambay@oac.unc.edu.ar}

\begin{abstract}
  We study the horseshoe dynamics of a low-mass planet in a
  three-dimensional, globally isothermal, inviscid disk. We find, as
  reported in previous work, that the boundaries of the horseshoe
  region (separatrix sheets) have cylindrical symmetry about the
  disk's rotation axis. We interpret this feature as arising from the
  fact that the whole separatrix sheets have a unique value of
  Bernoulli's constant, and that this constant does not depend on
  altitude, but only on the cylindrical radius, in barotropic
  disks. We next derive an expression for the torque exerted by the
  horseshoe region onto the planet, or horseshoe drag. Potential
  vorticity is not materially conserved as in two-dimensional flows,
  but it obeys a slightly more general conservation law (Ertel's
  theorem) which allows to obtain an expression for the horseshoe drag
  identical to the expression in a two-dimensional disk. Our results
  are illustrated and validated by three-dimensional numerical
  simulations. The horseshoe region is found to be slightly more
  narrow than previously extrapolated from two-dimensional analyses
  with a suitable softening length of the potential. We discuss the
  implications of our results for the saturation of the corotation
  torque, and the possible connection to the flow at the Bondi scale,
  which the present analysis does not resolve.
\end{abstract}


\section{Introduction}
\label{sec: Intro}
The tide between a low-mass protoplanet and a gaseous protoplanetary
disk features two components: the Lindblad torque, which arises from
the spiral wake that the embedded planet excites in the disk
\citep{ww86,og2002}, and the corotation torque, which comes from
material slowly drifting near the orbit. In the linear regime, the
corotation torque in two-dimensional disks scales with the disk's
vortensity\footnote{The ratio of the vertical component of the flow's
  vorticity to the surface density.}  gradient \citep[][hereafter
TTW02]{tanaka2002}.  In inviscid, two-dimensional disks, the
corotation torque has been found to always become non-linear,
regardless of the planet mass \citep{2009arXiv0901.2265P}, and to
coincide in this regime with the horseshoe drag~\citep{wlpi91}. In
two-dimensional barotropic disks, the magnitude of the horseshoe drag
scales also with the vortensity gradient \citep{cm09}.  This
dependence is a consequence of the material conservation of vortensity
in an inviscid, two-dimensional, barotropic flow.

Much less is known of the corotation torque in three-dimensional
disks. The linear corotation torque has been derived semi-analytically
by TTW02 in three-dimensional, globally isothermal disks. They find
that it nearly scales with the vortensity gradient. However, no
horseshoe drag expression has yet been established in three
dimensions, and so far it is unknown whether it would scale with the
vortensity gradient.

The purpose of this work is to provide a first step toward a better
understanding of the horseshoe dynamics in three dimensions, and the
torque exerted by the horseshoe region on the planet. In a recent
work, \citet{2015arXiv150503152F} have found that the horseshoe flow
around a nearly thermal mass planet is columnar: the shape of the
outer and inner boundaries of the horseshoe region is nearly
cylindrical, and the width of the horseshoe region barely depends on
altitude. \citet{2015MNRAS.452.1717L}, using numerical simulations
that include sophisticated physics (such as radiative transfer and
irradiation of the disk photosphere), have found that the horseshoe
region is marginally narrower (by $\sim 10$~\%) than in
two-dimensional setups with a nominal value of the softening length of
the potential. We will compare these findings to our results.

We follow hereafter an approach inspired in \citet{mak2006} and
\citet{cm09}: we use Bernoulli's invariant to provide useful
information about the horseshoe region and to derive a horseshoe drag
expression.  We limit ourselves here to inviscid, globally isothermal
disks, and we focus on the unsaturated horseshoe drag: we assume that
a sufficient amount of time has elapsed after the planet insertion for
the corotation torque to become non-linear, but we also assume that
this time is shorter than the time that separates two consecutive
close encounters of a fluid element with the planet, which is usually
much longer than the U-turn time scale. While \citet{mak2006}
investigated the transition between low and large mass planets, and
found a boost of the horseshoe width (and drag) for intermediate mass,
non gap-opening planets, we restrict ourselves here to the low-mass
case, for which the width of the horseshoe region scales as the square
root of the planetary mass.

Our paper is organized as follows: in section~\ref{sec: Eqs} we define
our notation and give the governing equations of the problem. In
section~\ref{sec:unsat-hors-drag}, we discuss some topological
properties of the horseshoe region, and we derive a horseshoe drag
expression valid exclusively in globally isothermal disks, using
Ertel's potential vorticity theorem \citep{ertel42}. This derivation
allows us to give an adequate definition of the vortensity gradient in
three-dimensional disks. In section~\ref{sec: Num}, we describe the
numerical code and the setup used in the numerical simulations that we
performed to check and illustrate the properties of
section~\ref{sec:unsat-hors-drag}. In sections~\ref{sec:summary}
and~\ref{sec:discussion} we summarize and discuss our results.

\section{Notation \& Equations}
\label{sec: Eqs}
\subsection{Notation}
We study the tidal interaction between a gaseous globally isothermal
(hence barotropic) protoplanetary disk in rotational equilibrium
around a star of mass $M_\star$, and an embedded protoplanet of mass
$M_p=qM_\star$, on a fixed circular orbit in the disk's midplane, with
orbital radius $r_p$ and orbital frequency $\Omega_p$.  We give the
position of a given fluid element either in spherical coordinates
$(r,\theta ,\phi)$ in a frame corotating with the planet, or in
cylindrical coordinates, $R = r \sin\theta$ being the cylindrical
radius and $z = r\cos\theta$ being the altitude. In each case the
frame is centered on the star and such that the planet's
$\phi$-coordinate equal zero. The angular speed of a fluid element in
a non-rotating frame is denoted $\Omega(r,\theta,\phi)$. We denote
with a $0$ subscript a quantity in the unperturbed disk ({\em i.e.}
prior to the insertion of the planet), and with a $p$ subscript a
quantity evaluated at $r=r_p$.  The surface density is defined as
\begin{equation}
  \label{eq:1}
  \Sigma(R)=\int_ {-\infty}^{+\infty}\rho(R,z)dz,
\end{equation}
where $\rho$ is the volumic density. In the unperturbed disk, its midplane value follows a power law:
\begin{equation}
  \label{eq:2}
 \rho(r,\theta=\pi/2,\phi)\propto r^{-\xi}. 
\end{equation}
The (uniform) soundspeed is denoted $c_s$, the pressure scale height
is $H(r)=c_s\sqrt{r^3/GM_\star}$, and the disk aspect ratio is
$h(r)=H(r)/r$.  Since the disk is globally isothermal, the aspect
ratio increases as the square root of the radius:
$h(r) = h_p(r/r_p)^{1/2}$.  One can work out the exact dependence of
the volumic density and rotational velocity on $r$ and $\theta$, for
the case in which the temperature is a power law of the spherical
radius, as described in appendix~\ref{sec:prof-disk-rotat}. From these
dependencies, one can infer that the surface density also obeys a
power law of radius: $\Sigma\propto r^{-\alpha}$, where $\alpha$ is
given by Eq.~(\ref{eq:88}).

We introduce the vorticity of the flow $\boldsymbol\zeta$ viewed in an inertial frame:
\begin{equation}
  \label{eq:3}
  \boldsymbol\zeta = \boldsymbol\nabla\times\boldsymbol v+2\boldsymbol\Omega_p,
\end{equation}
where $\boldsymbol v$ is the linear velocity in the frame corotating with the planet.
Throughout this work, we denote the potential vorticity (PV) with:
\begin{equation}
\label{eq:4}
\boldsymbol{w} = \frac{\boldsymbol{\zeta}}{\rho}.
\end{equation}
In order to use a vocabulary consistent with prior work in two dimensions, we
reserve the name vortensity for an adequate vertical integral of the
PV, which we shall introduce later.

We define the corotation sheet as the two-dimensional region where
$\Omega_0(r,\theta,\phi) = \Omega_p$. This sheet intersects the midplane
at $r=r_c(\phi)$.

We focus on the horseshoe region, defined by the particles that cross
over from the inner to the outer disk, or vice-versa, after a close
encounter with the planet. We call the boundary between the horseshoe
region and the rest of the disk the separatrix sheet. The streamlines
belonging to this sheet are called either the separatrix streamlines,
critical streamlines, or widest horseshoes
\citep{2015arXiv150503152F}.

At large azimuthal elongation from the planet, the horseshoe region can be divided
into four regions. The front (rear) part corresponds to fluid elements with
$\phi > 0$ ($\phi < 0$). Inside the front and rear parts, we distinguish the
upstream and downstream regions. The upstream region is the set of fluid
elements that have not experienced yet a close encounter with the planet,
whereas the downstream region is the set of fluid elements that have already
experienced a close encounter, and therefore crossed the corotation by
definition of the horseshoe region. In the front part, the upstream region
corresponds to $r>r_c$, and the downstream region to $r<r_c$. Opposite relations
hold in the rear part.

\subsection{Governing equations}
\label{sec:governing-equations}
The equations that govern the evolution of the flow are the continuity equation
and the Euler equations, which read respectively, in cylindrical coordinates and
in the frame corotating with the planet:
\begin{equation}
  \label{eq:5}
  \partial_t\rho+\frac{1}{R}\partial_R(R\rho v_R)+\frac
  1R\partial_\phi(\rho v_\phi)+\partial_z(\rho v_z)=0,
\end{equation}
\begin{multline}
 \label{eq:6}
 \partial_tv_R+v_R\partial_Rv_R+\frac{v_\phi}{R}\partial_\phi
 v_R+v_z\partial_z v_R\\-R\Omega_p^2-2\Omega_pv_\phi-\frac{v_\phi^2}{R}
 =-\frac{\partial_RP}{\rho}-\partial_R\Phi,
\end{multline}
\begin{equation}
 \label{eq:7}
 \partial_t j+v_R\partial_Rj+\frac{v_\phi}{R}\partial_\phi
 j+v_z\partial_z j=-\frac{\partial_\phi P}{\rho}-\partial_\phi\Phi,
\end{equation} 
and
\begin{equation}
 \label{eq:8}
 \partial_t v_z+v_R\partial_Rv_z+\frac{v_\phi}{R}\partial_\phi
 v_z+v_z\partial_zv_z=-\frac{\partial_zP}{\rho}-\partial_z\Phi,
\end{equation} 
where $v_R$, $v_\phi$ and $v_z$ are the cylindrical components of the
velocity in the rotating frame, $P=\rho c_s^2$ is the pressure,
$j=R^2\Omega=R^2\Omega_p+Rv_\phi$ is the specific angular momentum,
evaluated in the non-rotating frame, and $\Phi$ is
the gravitational potential given by:
\begin{multline}
  \label{eq:9}
  \Phi=-\frac{GM_\star}{\sqrt{R^2+z^2}}-\frac{GM_p}{\sqrt{R^2+r_p^2-2Rr_p\cos\phi+z^2}}\\+\frac{GM_p}{r_p^2}R\cos\phi,
\end{multline}
where the different terms are respectively the star potential, the
planet potential and the indirect term arising from the acceleration
of our non-inertial frame, centered on the star.

\section{Unsaturated horseshoe drag}
\label{sec:unsat-hors-drag}
We make the assumption that the flow in the vicinity of the planet is
in steady state, and that the upstream regions are essentially
unperturbed. As has been discussed previously
\citep{cm09,mc09,pbck10}, this amounts to considering the flow at a
time after the insertion of the planet that is larger than the U-turn
timescale, but shorter than (half) the horseshoe libration timescale,
in order to avoid possible saturation effects. The torque given by our
analysis is therefore the unsaturated horseshoe drag: the material
that experiences close encounters with the planet is ``fresh''
material, which has never experienced another close encounter
previously.

\subsection{A Bernoulli invariant}
\label{sec:bernoulli-invariant}
The right hand side of Eqs.~(\ref{eq:6})
to~(\ref{eq:8}) can be recast as the gradient of the effective potential
\begin{equation}
  \label{eq:10}
  \tilde \Phi=\Phi+\eta,
\end{equation}
where 
\begin{equation}
  \label{eq:11}
  \eta = c_s^2\log \left(\frac{\rho}{\rho_{00}}\right),
\end{equation}
is the fluid enthalpy, and $\rho_{00}$ is an arbitrary constant
dimensionally homogeneous to a density. Like the gravitational
potential, the enthalpy is defined to within an additive constant, and
specifying $\rho_{00}$ amounts to choosing for which value of the
density the enthalpy vanishes. In what follows we set
$\rho_{00}=M_*r_p^{-3}$. Multiplying Eqs.~(\ref{eq:6}), (\ref{eq:7})
and~(\ref{eq:8}) respectively by $v_R$, $v_\phi/R$ and $v_z$, and
summing the results under the assumption of a steady flow, we are left
with:
\begin{equation}
  \label{eq:12}
  \boldsymbol v\cdot\boldsymbol\nabla (E_{\rm kin}+\tilde\Phi)-Rv_R\Omega_p^2=0,
\end{equation}
where $\boldsymbol v\cdot\boldsymbol\nabla  \equiv v_R\partial_R
+\frac{v_\phi}{R}\partial_\phi+v_z\partial_z$ and 
\begin{equation}
  \label{eq:13}
  E_{\rm kin}=\frac 12(v_R^2+v_\phi^2+v_z^2),
\end{equation}
 hence we have:
\begin{equation}
  \label{eq:14}
  \boldsymbol v\cdot\boldsymbol\nabla B_J=0,
\end{equation}
where
\begin{equation}
  \label{eq:15}
  B_J=E_{\rm kin}+\tilde\Phi-\frac 12R^2\Omega_p^2,
\end{equation}
is a Bernoulli invariant. The $J$ index is in analogy with the Jacobi
constant of a test particle, and is also meant to avoid confusion with
Oort's second constant, that we shall introduce later. This invariant
is conserved along the streamlines of the domains over which
Eq.~(\ref{eq:6}) to~(\ref{eq:8}) hold, that is to say over the domains
that do not contain shocks. We assume this to be the case, in
particular, of the coorbital region of sufficiently low mass planets.

\subsection{Considerations about the topology of the horseshoe region}
\label{sec:cons-about-topol}
In previous two-dimensional analyses \citep{cm09,mc09}, the width of
the horseshoe region far from the planet can be determined using the
fact that Bernoulli's invariant is the same at the stagnation point and
far away on the separatrices. From Eq.~(\ref{eq:15}) this value of the
Bernoulli invariant depends exclusively on the enthalpy of the fluid at
the stagnation point and its location. We discuss below whether a
similar property can be generalized to the three-dimensional case.

The first question to address is therefore that of the set of stagnation
points. A stagnation point, in three dimensions, is found wherever the three
components of the velocity simultaneously vanish. In general, the constraint
$v_\phi=0$ is verified on a two-dimensional manifold (the corotation sheet),
within which the additional constraint $v_R=0$ yields a one-dimensional
manifold. Then, in general, the additional constraint $v_z=0$ yields a finite
number of points within this one-dimensional manifold. We therefore expect to
have a finite number of stagnation points in three dimensions.  The streamline
analysis performed on data of three-dimensional calculations, presented in
section~\ref{sec:separ-sheet-stagn}, corroborates this statement. 

In the following we shall assume that in the three-dimensional case
the horseshoe region is still bounded by a well-defined set of
streamlines, the separatrix sheets, three-dimensional generalization
of the separatrix streamlines of the two-dimensional cases. We will
come back to this assumption in the discussion of
section~\ref{sec:relat-with-flow}.  The second question is whether all
fluid elements of a separatrix sheet go through a stagnation
point. While the answer to this question is trivial in the
two-dimensional case, it is not that obvious in the three-dimensional
case. In particular, one can imagine that a given fluid element of a
separatrix, originating at a given altitude, will complete a horseshoe
U-turn without ever reaching the altitude of the stagnation
point. Assume that such a fluid element exists. By hypothesis, the
norm of its velocity is therefore finite everywhere on its associated
streamline\footnote{This streamline coincides with the path followed
  by the fluid element, under our assumption of a steady state.}. Then
consider another fluid element on a neighboring streamline, initially
close to the first one. The separation $\boldsymbol\xi$ between both fluid
elements then obeys $d\boldsymbol\xi/dt = \boldsymbol v_2-\boldsymbol v_1$, where
$\boldsymbol v_2$ and $\boldsymbol v_1$ are the velocities of both fluid elements
along their trajectories. One can recast this derivative in terms of
the curvilinear abscissa $s$ along the trajectory of the first fluid
element: $d\boldsymbol\xi/ds=(\boldsymbol v_2-\boldsymbol v_1)/v_1$. Since $v_1$ remains
finite and since the velocity field is continuous, $d\boldsymbol\xi/ds$ can
be made arbitrarily small provided the fluid elements are sufficiently
close initially. Upon integration over any finite length $s$, this
means that two neighboring streamlines can always remain arbitrarily
close to each other, provided their initial separation is adequately
chosen. This is in contradiction with the character of a separatrix:
two neighboring streamlines on opposite sides of the separatrix sheet
always follow divergent paths after a finite length, no matter how
close they are initially. This shows that the modulus of the velocity
vector along a separatrix streamline must vanish somewhere, which
implies the passage through a stagnation point. There is an 
immediate consequence to this statement:
the streamlines of the separatrix sheets have significant vertical excursions in
order to pass through one of the stagnation points. By continuity, the horseshoe
trajectories are vertically bent near the U-turns, especially those which are
close to the separatrix. This has already been observed in three-dimensional
calculations \citep{gda2003,2015arXiv150503152F}.

There is another consequence to the fact that all separatrices go
through at least one of the stagnation points: the value of
Bernoulli's invariant on any separatrix streamline must be the value
that it has at the corresponding stagnation point. By spawning
streamlines (downstream and upstream) in all directions in the
immediate vicinity of a stagnation point, one generates a separatrix
sheet with a unique value of Bernoulli's invariant.

As in the two-dimensional case, there may be several stagnation
points, each with its own value of the Bernoulli invariant. Each of
these points generates a separatrix sheet where Bernoulli's invariant
has everywhere the same value as at the point. We note that two
separatrix sheets with different values of Bernoulli's invariant
cannot cross, so that they are disjoint or nested. We also show in
appendix~\ref{sec:uniq-bern-invar} that a given separatrix sheet
cannot be connected to two stagnation points with different values of
Bernoulli's invariant.

In what follows, we are interested
in the separatrix sheet that lies furthest from the orbit,
corresponding to the widest horseshoe streamlines.  This is to be
compared to the two-dimensional situation, which can feature two
X-stagnation points \citep{cm09, 2013MNRAS.430.1764G}, but the one
that determines the overall horseshoe dynamics is the one that has the
lowest value of the Bernoulli invariant (hence connected to
streamlines that lie the furthest from the orbit).

An important hypothesis is that the flow is barotropic: all variables
of Eq.~(\ref{eq:15}) are continuous in a barotropic fluid, hence the
value of Bernoulli's invariant is well-defined in the vicinity of a
stagnation point. In a baroclinic situation (if, for instance, the
flow obeyed an energy equation), the enthalpy, hence Bernoulli's
invariant, would not necessarily be continuous.

The fact that Bernoulli's invariant is uniform on a separatrix sheet
provides an idea of the vertical shape of the horseshoe region far
from the planet. Considering that the flow at large azimuthal distance
from the planet is unperturbed, we can simplify the expression of
Bernoulli's invariant given by Eq.~(\ref{eq:15}) as:
\begin{equation}
  \label{eq:16}
  B_J = \frac 12v_\phi^2+\Phi_*-\frac
  12R^2\Omega_p^2+c_s^2\log\left(\frac{\rho_0}{\rho_{00}}\right),
\end{equation}
where $\Phi_*(\boldsymbol r)$ is the stellar potential. As shown in
appendix~\ref{sec:prof-disk-rotat}, the azimuthal velocity in a globally
isothermal disk in rotational and hydrostatic equilibrium does not depend on the
altitude (Eq.~\ref{eq:84}), so the vertical derivative of Bernoulli's invariant
reduces to:
\begin{equation}
  \label{eq:17}
  \partial_zB_J=\partial_z\Phi_*+c_s^2\partial_z\log\left(\frac{\rho_0}{\rho_{00}}\right).
\end{equation}
The right hand side of Eq.~(\ref{eq:17}) cancels out, as can be seen
from Eq.~(\ref{eq:8}) for an axisymmetric disk in equilibrium, hence
Bernoulli's invariant in the unperturbed disk depends only on the
cylindrical radius. The horseshoe's separatrix sheets should
therefore be cylinders coaxial with the disk's rotation axis, and so
should the corotation sheet, since $v_\phi$ also depends exclusively
on the cylindrical radius. The horseshoe region is therefore expected
to have a constant width with altitude. This has been seen by
\citet{2015arXiv150503152F}, who describe the horseshoe flow as
columnar.  This property will be corroborated by numerical simulations
in section~\ref{sec: Num}. This allows us to introduce unambiguously
the (unique) half-width of the horseshoe region, which we denote $x_s$
as in previous two-dimensional work.

Note that the cylindrical shape of the corotation sheet is a consequence of
the disk being globally isothermal. Expanding in $z$ the relations
given in appendix~\ref{sec:prof-disk-rotat}, we can see that for the
case in which the temperature increases outward, the corotation has
its smallest radius at the midplane, and the opposite holds for the
more realistic case corresponding to a temperature decreasing outward:
its largest radius is at the midplane.

\subsection{Torque expression}
\label{sec:torque-expression}
\subsubsection{Domain of interest}
\label{sec:domain-interest}
As in \citet{cm09} and \citet{mc09}, we define a domain of interest ${\cal D}$
that encloses the planet and the horseshoe region, that extends from $z=-\infty$
to $z=+\infty$, from $\phi=\phi_-$ to $\phi=\phi_+$, where $\phi_-< 0$ and
$\phi_+>0$, and where $|\phi_-|$ and $\phi_+$ are chosen sufficiently large so
that the planetary potential is negligible at $\phi=\phi_\pm$. Similarly, we
chose the radial boundaries of the domain, $R_\mathrm{min}$ and
$R_\mathrm{max}$, to cover an extent much wider than the horseshoe region. Our
domain therefore encloses not only horseshoe streamlines, but also streamlines
that do not perform a U-turn, and which belong to the inner or outer disk. We
make the assumption that the horseshoe flow reaching $\phi_\pm$ is relaxed in
the sense that the flow velocity is exclusively azimuthal, and the azimuthal
derivatives of the hydrodynamic variables vanish. Naturally, a given fluid
element starting its horseshoe trajectory at a given altitude $z_0$ may change
its altitude during its journey to the downstream part of the horseshoe flow. As
we shall see in numerical simulations, it may also oscillate vertically a few
times after the U-turn. We make no restrictive assumption concerning its final
altitude, which may be different from $z_0$. Our only assumption at this stage
is that its altitude has converged to a constant value by the time it reaches
the azimuth $\phi_\pm$, which is why we must consider sufficiently large values
of $|\phi_\pm|$.

The torque exerted by the material enclosed in the domain ${\cal D}$
on the planet is:
\begin{equation}
  \label{eq:18}
  \Gamma = \int\!\!\int\!\!\int_{\cal D}\rho\partial_\phi\Phi RdRdzd\phi.
\end{equation}
Using Eqs.~(\ref{eq:7}), then~(\ref{eq:5}), under our assumption of steady
state, this integral can be recast as integrals that account for the angular
momentum budget on the faces of the domain:
\begin{eqnarray}
  \Gamma&=&-\left.\int_{-\infty}^{+\infty}\!\!\!\!dz\int_{\phi_-}^{\phi_+}\!\!\!\!d\phi\,\rho v_R j
  R\right|_{R_\mathrm{max}}
+\left.\int_{-\infty}^{+\infty}\!\!\!\!dz\int_{\phi_-}^{\phi_+}\!\!\!\!d\phi\,\rho v_R j
  R\right|_{R_\mathrm{min}}\nonumber\\
&&-\left.\int_{-\infty}^{+\infty}\!\!\!\!dz\int_{R_\mathrm{min}}^{R_\mathrm{max}}\!\!\!\!dR\,(\rho
   v_\phi j+RP)\right|_{\phi_+}\nonumber\\
&&+\left.\int_{-\infty}^{+\infty}\!\!\!\!dz\int_{R_\mathrm{min}}^{R_\mathrm{max}}\!\!\!\!dR\,(\rho v_\phi j+RP)\right|_{\phi_-}.
  \label{eq:19}
\end{eqnarray}
The edge terms in $z$ cancel out since the density vanishes at larger
altitude. The integrals of Eq.~(\ref{eq:19}) involve angular momentum fluxes on
the arc-of-cylinder faces of the domain (in $R=R_\mathrm{min}$ and $R=R_\mathrm{max}$),
and on the radial faces of the domain (in $\phi=\phi_\pm$).

\subsubsection{Corotation torque integral}
\label{sec:corot-torq-integr}
We now assume that the angular momentum flux on the arc-of-cylinder faces is
the flux carried by the spiral wake, which corresponds to the Lindblad torque on
the planet, and that the fluxes on the radial faces of the domain correspond, on
the contrary, to the corotation torque. We assume that there is a range of
aspect ratios $(R_\mathrm{max}-R_\mathrm{min})/[r_p(\phi_+-\phi_-)]$
of the domain for which
this separation of the torques is correct, and we focus exclusively on the flux
on the radial faces, which we interpret as the full, non-linear corotation
torque:

\begin{multline}
  \label{eq:20}
  \Gamma_\mathrm{CR}=-\left.\int_{-\infty}^{+\infty}\!\!\!\!dz\int_{R_\mathrm{min}}^{R_\mathrm{max}}\!\!\!\!dR\,(\rho
   v_\phi j+RP)\right|_{\phi_+}\\
+\left.\int_{-\infty}^{+\infty}\!\!\!\!dz\int_{R_\mathrm{min}}^{R_\mathrm{max}}\!\!\!\!dR\,(\rho v_\phi j+RP)\right|_{\phi_-}.
\end{multline}

We lay down a last assumption: there is no shock in the domain of
interest $\mathcal{D}$. As the latter includes the planetary wake in
the vicinity of the planet, this domain should be narrow enough so as
not to reach the location at which the wake eventually shocks as a
consequence of wave steepening. This location can be dangerously close
to the planet owing to the differential rotation of the disk
\citep{gr2001}, so we focus on planetary masses largely below the
thermal mass. This assumption is required, \emph{stricto sensu}, to
satisfy our steady state assumption. Should shocks be present over the
domain, they would transfer the planetary torque to the disk material
and they would carve a gap~\citep{rafikov02} over a time scale longer
than the typical time considered here, intermediate between the
horseshoe U-turn and libration time scales, precluding the existence
of a steady state.  Arguably shocks could still appear beyond the
radial limits of the domain $\mathcal{D}$. We assume that radial
boundary conditions are tailored to ensure a steady state over
$\mathcal{D}$.

Using Eq.~(\ref{eq:16}) and the assumption that the motion is exclusively
azimuthal at large azimuthal distances from the planet, we write:
\begin{equation}
  \label{eq:21}
  \partial_RB_J = \zeta_zv_\phi,
\end{equation}
and we change our integration variable to $B_J$ so as to transform Eq.~(\ref{eq:20}) into:
\begin{eqnarray}
  \Gamma_\mathrm{CR}&=&-\left.\int_{-\infty}^{+\infty}\!\!\!\!dz\int_{B_m}^{B_c}\!\!\!\!dB_J\frac{\rho
   v_\phi
                        j+RP}{\zeta_zv_\phi}\right|_{\phi_+,R<R_c}\nonumber\\
&&-\left.\int_{-\infty}^{+\infty}\!\!\!\!dz\int_{B_m}^{B_c}\!\!\!\!dB_J\frac{\rho v_\phi j+RP}{\zeta_zv_\phi}\right|_{\phi_+,R>R_c}\nonumber\\
&&+\left.\int_{-\infty}^{+\infty}\!\!\!\!dz\int_{B_m}^{B_c}\!\!\!\!dB_J\frac{\rho
   v_\phi
                j+RP}{\zeta_zv_\phi}\right|_{\phi_-,R<R_c}\nonumber\\
&&+\left.\int_{-\infty}^{+\infty}\!\!\!\!dz\int_{B_m}^{B_c}\!\!\!\!dB_J\frac{\rho v_\phi j+RP}{\zeta_zv_\phi}\right|_{\phi_-,R>R_c},
  \label{eq:22}
\end{eqnarray}
where $B_m$ represents the minimum value of Bernoulli's invariant, at
larger distance from corotation (without loss of generality we can
assume that it has same value at the outer and inner sides), $B_c$
represents the value of Bernoulli's invariant at corotation, and $B_s$
its value on the separatrix sheet.

Note that the relation between $R$ and $B_J$ is not
one-to-one. Eq.~(\ref{eq:21}) shows that the radial derivative of
Bernoulli's invariant changes sign at corotation, where $B_J$ is
maximum. We therefore had to split in two each term of
Eq.~(\ref{eq:20}), specifying whether we consider contributions from
inside or outside corotation. We number one to four, in their order of
appearance, the four terms of the right hand side of
Eq.~(\ref{eq:22}). Terms one and four correspond to downstream
material leaving the domain, while terms two and thee correspond to
upstream material entering it. Regardless of the complexity of the
flow in the vicinity of the planet, a fluid element entering the
domain either leaves it on the same side of corotation, or on the
other side.

\begin{figure}
  \includegraphics[width=1.05\columnwidth]{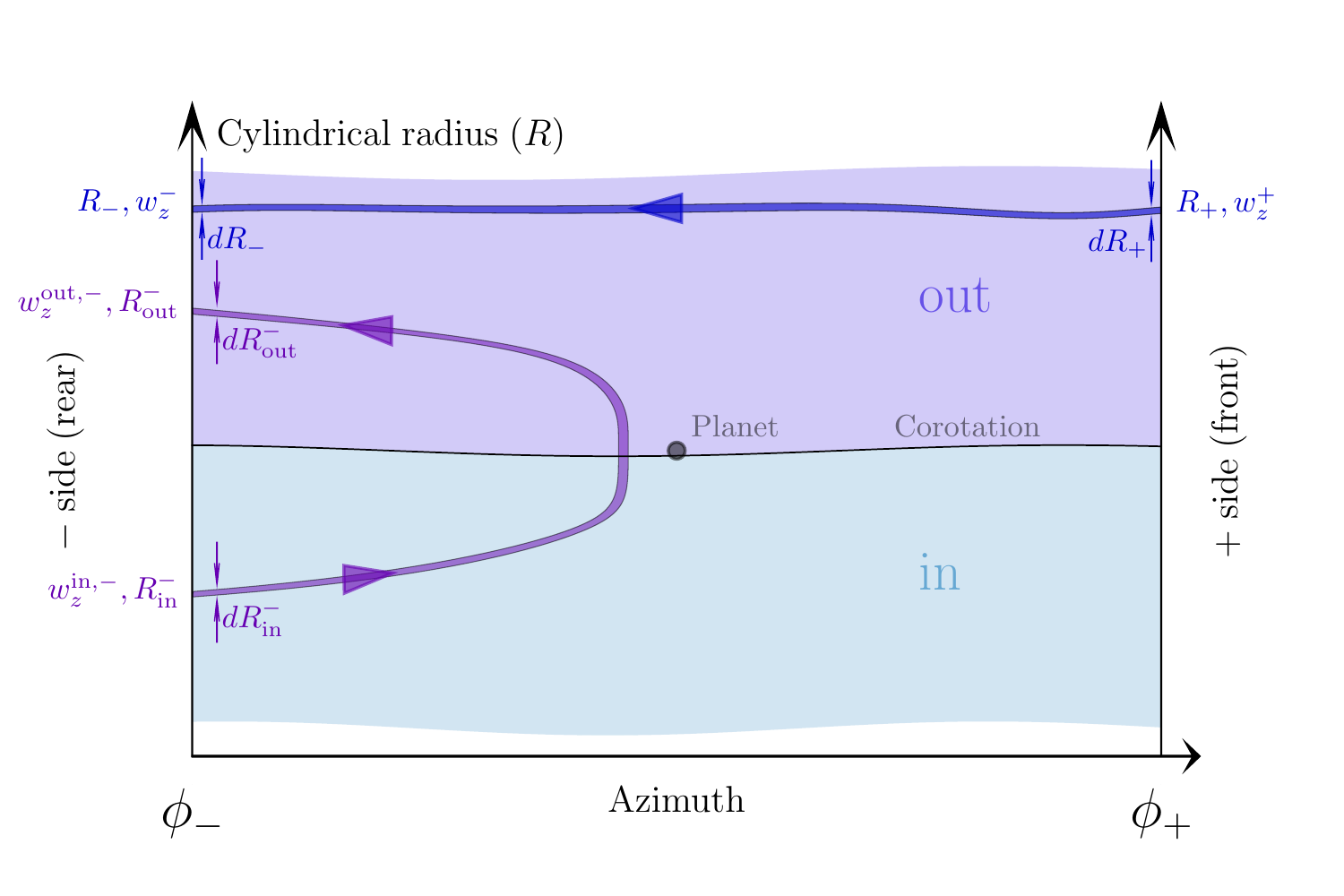}
  \caption{Sketch of circulating (blue) and librating (purple) stream
    tubes. Each tube is associated to a given range $[B_J,B_J+dB_J]$
    of Bernoulli's invariant (different in the two cases). For the sake
    of legibility, not all the variables used in the derivation in the
    text are shown, but their meaning and the location at which they
    are defined are straightforward to infer from the variables shown
    here. For the same reason, the vertical dimension is not
    represented on this sketch, which shows a projection on the
    horizontal plane $(R,\phi)$. It must be kept in mind that the
    endpoints of the stream tubes may have different altitudes, and
    different vertical extents.}
  \label{fig:sktech}
\end{figure}

We examine firstly the case of a fluid element leaving the domain on
the same side of corotation (``circulating'' rather than
``librating''). For the sake of definiteness we assume
it enters the domain in $\phi_+$, $R>r_c$. It therefore leaves it in
$\phi_-$, $R>r_c$. An example of the stream tube defined by such fluid
element is depicted in blue in Fig.~\ref{fig:sktech}. The
contribution of this term to the second integral of Eq.~(\ref{eq:22})
is:
\begin{equation}
  \label{eq:23}
  d\Gamma^+_\mathrm{CR}=-\left.dB_Jdz_+\frac{\rho v_\phi j+RP}{\zeta_zv_\phi}\right|_{\phi_+}
\end{equation} 
and its contribution to the fourth one is:
\begin{equation}
  \label{eq:24}
  d\Gamma^-_\mathrm{CR}=\left.dB_Jdz_-\frac{\rho v_\phi j+RP}{\zeta_zv_\phi}\right|_{\phi_-}.
\end{equation}
Since we assumed there are no shocks on the domain, the fluid elements have
their Bernoulli's invariant materially conserved. We can work out the net flux
contribution by summing  Eqs.~(\ref{eq:23}) and~(\ref{eq:24}).

The fluid element may exit the domain at a radius and altitude
different from those at which it entered the domain, and its velocity
and enthalpy may also be different. We assume the variations of
radius, velocity and enthalpy to be sufficiently small to allow first
order expansions. We denote with $\delta X$ the small variation of a
quantity $X$ at the fluid element location, between its arrival to and
exit from the domain (\emph{i.e.} the Lagrangian variation seen by the
fluid element), and with $\delta'X$ the variation of that quantity
between the two radial faces, at given radius and altitude. Expanding
Eq.~(\ref{eq:16}), making use of Eqs.~(\ref{eq:6}) and~(\ref{eq:8})
together with the assumptions we laid down at large azimuthal distance
from the planet ($v_R=v_z=0$, $\partial_\phi\equiv 0$), we are led to:
\begin{equation}
  \label{eq:25}
  \delta B_J = \left(\frac{v_\phi}{R}\right)_{\phi_-}\delta j+\delta'\eta,
\end{equation}
with $\delta j=j_--j_+$ and $\delta'\eta = \eta(R,z,\phi_-)-\eta(R,z,\phi_+)$.
Since $\delta B_J=0$, we have:
\begin{eqnarray}
  d\Gamma^-_\mathrm{CR} &=& \frac{dB_Jdz_-}{\zeta_z^-}\left[\rho_-
 \left(j_+-\frac{R\delta'\eta}{v_\phi}\right)
                            +\frac{RP}{v_\phi}\right]_{\phi_-}\nonumber\\
&=&\frac{dB_Jdz_-}{w_z^-}j_++dR_-dz_-R_-P(R_-,z_-,\phi_+),
  \label{eq:26}
\end{eqnarray}
where we have used the first order expansion $\delta'P=\rho_-\delta'\eta$, and
where $dR_\pm$ is the radial width of the stream tube at azimuth $\phi_\pm$. In
order to further transform the first term, we use Ertel's theorem\footnote{The
  original paper of Ertel has been translated to English by
  \citet{erttrans2004}.} \citep{ertel42}. Ertel's theorem, in a reduced form
that is of interest here, states that in a barotropic flow we have the
conservation law:
\begin{equation}
  \label{eq:27}
  D_t\left(\frac{\boldsymbol\zeta}{\rho}\cdot \boldsymbol{\nabla}\psi \right) = 0,
\end{equation}
where $\psi$ is any quantity materially conserved by the flow
($D_t\psi=0$).  We can take here for $\psi$ an arbitrary function
whose derivative in $z$ does not vanish, such as $z_+$ itself,
regarded as a Lagrangian, passive scalar on the fluid
elements. Ertel's theorem reads in this case:
\begin{equation}
  \label{eq:28}
  \left(\boldsymbol w\cdot\boldsymbol{\nabla}z_+\right)_{\phi_+}=\left(\boldsymbol w\cdot\boldsymbol{\nabla}z_+\right)_{\phi_-},
\end{equation}
or
\begin{equation}
  \label{eq:29}
  w_z^+ = w_z^-\frac{dz_+}{dz_-}+w_R^-\frac{dz_+}{dR}.
\end{equation}
Note that in Eq.~(\ref{eq:29}) there is no azimuthal derivative, as per our
assumption that these vanish at large azimuthal distance from the
planet. The factor $w_R^-$ in the last term is expected to be much
smaller than $w_z^\pm$ (we will see in
section~\ref{sec: Num} that the vorticity tilt angle, especially in the circulating
region, is well below $10^{-2}$), and we expect the second factor
of this term ($dz_+/dR$, which represents the inclination of an
initially horizontal sheet of material after the interaction with the
planet), to be  also a small number.
We therefore write:
\begin{equation}
  \label{eq:30}
  \frac{dz_+}{w_z^+}\approx \frac{dz_-}{w_z^-},
\end{equation}
with an accuracy largely better than the $10^{-2}$ level. The net contribution to the
corotation torque of the stream tube that we considered is therefore:
\begin{multline}
  \label{eq:31}
  d\Gamma_\mathrm{CR}=d\Gamma_\mathrm{CR}^++d\Gamma_\mathrm{CR}^-=dR_-dz_-R_-P(R_-,z_-,\phi_+)\\-dR_+dz_+R_+P(R_+,z_+,\phi_+).
\end{multline}
We see that the torque contribution of our circulating fluid element only
amounts to the pressure on the \emph{upstream} side: both terms above
are evaluated in $\phi_+$ thanks to the transformation arising from
Eq.~(\ref{eq:25}). Anticipating on what follows, this pressure
contribution will cancel out upon the integration on all stream tubes,
so that the net contribution will exclusively arise from the horseshoe
region, which we now examine.

We consider a stream tube that executes a horseshoe
U-turn (corresponding to ``librating'' fluid elements). For the sake
of definiteness we assume it enters the domain of interest in $R<r_c$
and $\phi_-$ and exits it in $R>r_c$ and $\phi_-$, such as the one
depicted in purple in Fig.~\ref{fig:sktech}. Such stream tube yields
contributions to the third and fourth integrals of Eq.~(\ref{eq:22}),
which read respectively:
\begin{eqnarray}
  \label{eq:32}
 \left.d\Gamma_\mathrm{CR}^-\right|_\mathrm{in}&=&\left.dB_Jdz_-\frac{\rho v_\phi j+RP}{\zeta_zv_\phi}\right|_{\phi_-,\mathrm{in}}\\
  \label{eq:33}
 \left.d\Gamma_\mathrm{CR}^-\right|_\mathrm{out}&=&\left.dB_Jdz_-\frac{\rho v_\phi j+RP}{\zeta_zv_\phi}\right|_{\phi_-,\mathrm{out}},
\end{eqnarray}
where the subscript in or out specifies whether we consider a region inside or
outside corotation. In the above analysis of a circulating stream tube,
from Eq.~(\ref{eq:23}) to Eq.~(\ref{eq:31}), fluid elements did not cross
corotation, so there was no need to specify explicitly the location of each term
as they were all evaluated systematically outside corotation. As previously, we
can recast Eq.~(\ref{eq:32}) as:
\begin{equation}
  \label{eq:34}
  \left.d\Gamma_\mathrm{CR}^-\right|_\mathrm{in}=\frac{dB_Jdz^-_\mathrm{in}}{w_z^\mathrm{in,-}}j^-_\mathrm{in}-dR^-_\mathrm{in}dz^-_\mathrm{in}R^-_\mathrm{in}P(\phi_-,R^-_\mathrm{in},z^-_\mathrm{in}).
\end{equation}
We can write Eq.~(\ref{eq:33}) in a similar way, and transform it
using Eq.~(\ref{eq:25}), for which we set $\delta B_J=0$:
\begin{eqnarray}
  \left.d\Gamma_\mathrm{CR}^-\right|_\mathrm{out}&=&\!\frac{dB_Jdz^-_\mathrm{out}}{w_z^\mathrm{out,-}}\!j^-_\mathrm{out}\!+\!dR^-_\mathrm{out}dz^-_\mathrm{out}R^-_\mathrm{out}P(\phi_-,R_\mathrm{out},z_\mathrm{out})\nonumber\\
&=&\!\frac{dB_Jdz^-_\mathrm{out}}{w_z^\mathrm{out,-}}\!j^+_\mathrm{out}\!+\!dR^-_\mathrm{out}dz^-_\mathrm{out}R^-_\mathrm{out}P(\phi_+,R_\mathrm{out},z_\mathrm{out})
  \label{eq:35}
\end{eqnarray}
We see that this transformation allows to express the torque
contribution in term of the upstream pressure
$P(\phi_+,R^-_\mathrm{out},z^-_\mathrm{out})$ and specific angular
momentum $j^+_\mathrm{out}$ on the outer side. Since the
transformation of Eq.~(\ref{eq:35}) has been obtained by letting
$\delta B_J=0$, the quantity $j^+_\mathrm{out}$ represents the
specific angular momentum of the upstream material in the outer disk
that has same Bernoulli's invariant as our horseshoe stream tube.

When summing all contributions of Eq.~(\ref{eq:22}), it is now clear
that the contributions of the pressure torques cancel out, since they
all amount to summing the upstream pressure field: in the outer disk,
Eq.~(\ref{eq:31}) for the circulating fluid elements and
Eq.~(\ref{eq:35}) for the librating ones show that we only consider
the upstream pressure ($P(\phi_+)$). For the inner disk 
transformations similar to those considered previously would yield
torque contributions as function of $P(\phi_-)$, such as the
contribution given by Eq.~(\ref{eq:32}). Since the angular momentum
advective contributions cancel out for circulating streamlines, as shown by
Eq.~(\ref{eq:31}), we are only left with the contributions to the
angular momentum flux of the librating streamlines. For the rear
horseshoe, the total contribution therefore amounts to:
\begin{equation}
  \label{eq:36}
  \Gamma^-_\mathrm{CR}=\int_{-\infty}^{+\infty}dz\int_{B_s}^{B_c}\frac{dB_J}{w_z^\mathrm{in,-}}j^-_\mathrm{in}-\int_{-\infty}^{+\infty}dz\int_{B_s}^{B_c}\frac{dB_J}{w_z^\mathrm{out,-}}j^+_\mathrm{out}.
\end{equation}
We now exchange the integral sums, which yields:
\begin{equation}
  \label{eq:37}
  \Gamma^-_\mathrm{CR}=\int_{B_s}^{B_c}dB_J\int_{-\infty}^{+\infty}dz\left(\frac{j^-_\mathrm{in}}{w_z^\mathrm{in,-}}-
\frac{j^+_\mathrm{out}}{w_z^\mathrm{out,-}}\right)
\end{equation}
This expression exclusively features upstream values of the
specific angular momentum. As we consider the unsaturated horseshoe
drag, we assume that it has same value as in the unperturbed disk. It
is therefore independent of altitude, and can be taken out of the
integral on $z$:
\begin{equation}
  \label{eq:38}
  \Gamma^-_\mathrm{CR}=\int_{B_s}^{B_c}dB_J\left(
  j^-_\mathrm{in}\int_{-\infty}^{+\infty}\frac{dz}{w_z^\mathrm{in,-}}-
 j^+_\mathrm{out}\int_{-\infty}^{+\infty}\frac{dz}{w_z^\mathrm{out,-}}\right).
\end{equation}
We again apply Ertel's theorem with the ancillary function $\psi =
z^-_\mathrm{in}$, initial altitude of a fluid element, which gives:
\begin{equation}
  \label{eq:39}
  w_z^\mathrm{in,-}=w_z^\mathrm{out,-}\frac{\partial
    z^-_\mathrm{in}}{\partial
    z^-_\mathrm{out}}+w_R^\mathrm{out,-}\frac{\partial
    z^-_\mathrm{in}}{\partial R_\mathrm{out}}\approx w_z^\mathrm{out,-}\frac{\partial
    z^-_\mathrm{in}}{\partial
    z^-_\mathrm{out}},
\end{equation}
where we have again assumed that the radial tilt of the vortex tubes
is negligible. We can therefore write:
\begin{equation}
  \label{eq:40}
  \int_{-\infty}^{+\infty}\frac{dz_\mathrm{out}}{w_z^\mathrm{out,-}}=\int_{-\infty}^{+\infty}\frac{dz_\mathrm{out}}{w_z^\mathrm{in,-}}\frac{\partial
    z^-_\mathrm{in}}{\partial
    z^-_\mathrm{out}}=\int_{-\infty}^{+\infty}\frac{dz_\mathrm{in}}{w_z^\mathrm{in,-}}.
\end{equation}
We conclude that the integral $\int
w_z^{-1}dz$ has same value in the upstream and downstream flow, far from the
planet. The expression of this integral can be made slightly simpler
by making use of the inverse of the vertical component of the
PV, which we dubbed the load in an earlier paper
\citep{2010ApJ...723.1393M}:
\begin{equation}
  \label{eq:41}
  l = \frac{\rho}{\zeta_z},
\end{equation}
and we denote with $L$ the vertical integral of the load:
\begin{equation}
  \label{eq:42}
  L=\int_{-\infty}^{+\infty}l(z)dz.
\end{equation}
We can now transform Eq.~(\ref{eq:37}) into:
\begin{equation}
  \label{eq:43}
  \Gamma^-_\mathrm{CR}=\int_{B_s}^{B_c}dB_JL_\mathrm{in}(B_J)(j^-_\mathrm{in}-j^+_\mathrm{out}).
\end{equation}
The jump of specific angular momentum during a U-turn,
$(j^-_\mathrm{in}-j^+_\mathrm{out})$, is expressed in term of the
values in the unperturbed disk. It is a function of Bernoulli's
invariant exclusively, which we denote $-\Delta j_0$. The horseshoe
drag arising from the front part reads similarly:
\begin{equation}
  \label{eq:44}
  \Gamma^+_\mathrm{CR}=\int_{B_s}^{B_c}dB_JL_\mathrm{out}(B_J)\Delta
  j_0,
\end{equation}
and the net corotation torque is:
\begin{equation}
  \label{eq:45}
  \Gamma_\mathrm{CR}=\int_{B_s}^{B_c}dB_J(L_\mathrm{out}-L_\mathrm{in})\Delta j_0
\end{equation}
This integral, which only contains quantities of the unperturbed disk,
can be evaluated using Eq.~(\ref{eq:21}), which can be recast to
lowest order as $\partial_RB_J=4A_pB_px$ and gives the relationship:
\begin{equation}
  \label{eq:46}
  B_J=B_c+2A_pB_px^2,
\end{equation}
where $A=\frac 12 r\partial_r\Omega$  and
$B=\frac{1}{2r} \partial_rj_0$ are respectively Oort's first and
second constant.
Changing our variable of integration to $x$, we are led to:
\begin{equation}
  \label{eq:47}
  \Gamma_{\rm CR} = 8|A_p|B_p^2r_p\partial_RL_0x_s^4. 
\end{equation}
Defining as in earlier works the dimensionless vortensity gradient at the planet's 
location as:
\begin{equation}
  \label{eq:48}
  {\cal V} = r_p\frac{\partial_RL_0}{L_0},
\end{equation}
and assuming that the flow's vorticity barely depends on altitude in 
the unperturbed disk, so that $L_0\approx \Sigma_p/(2B_p)$, we are 
eventually led to:
\begin{equation}
  \label{eq:49}
  \Gamma_{\rm CR} = \frac 34\Sigma_p\Omega_p^2{\cal V}x_s^4,
\end{equation}
which is the exact same horseshoe drag expression as in the two 
dimensional case. 

We conclude that the horseshoe dynamics in a three-dimensional,
globally isothermal disk, despite the complexity of the three
dimensional flow, retain some of the simplicity of the two-dimensional
case, firstly because the horseshoe region width is independent of
altitude, which allows to use unambiguously a unique half width $x_s$,
and secondly because the drag expression is the same as in the
two-dimensional case, provided an adequate definition of the
vortensity is used. Eqs.~(\ref{eq:41}), (\ref{eq:42})
and~(\ref{eq:47}) show that the vortensity $V$ in a three-dimensional
disk must be defined as:
\begin{equation}
  \label{eq:50}
  V=\left[\int_{-\infty}^{+\infty}\left(\frac{\zeta_z}{\rho}\right)^{-1}dz\right]^{-1}.
\end{equation}

\section{Numerical simulations}
\label{sec: Num}
\subsection{Mesh geometry and set up}
\label{sec:mesh-geometry-set}
In order to illustrate and check the results of
section~\ref{sec:unsat-hors-drag}, we have undertaken numerical
simulations of a globally isothermal disk with a low mass planet
embedded on a fixed circular prograde orbit, coplanar with the
disk. We used for that purpose the public code FARGO3D\footnote{See
  \texttt{http://fargo.in2p3.fr}.} (Ben\'\i tez-Llambay \& Masset,
submitted), with the setup \texttt{p3diso}, meant to describe a
(locally or globally) isothermal gas in orbit around a point-like
mass, on a spherical mesh. The use of a spherical mesh may seem not
adapted to the problem at hand, since we expect from
section~\ref{sec:unsat-hors-drag} a cylindrical symmetry for a number
of variables, such as the azimuthal velocity or Bernoulli's
invariant. In a similar manner, the horseshoe separatrices are
expected to be cylinders. Our choice of a spherical mesh instead of a
cylindrical one fulfills two purposes:
\begin{itemize}
\item owing to the strong flaring of a globally isothermal disk
  ($H\propto R^{3/2}$), we avoid very empty regions at lower
  radius and high altitude. 
\item We can discard possible mesh effects if we find features with
  cylindrical symmetry.
\end{itemize}

Our mesh extends from $-\pi$ to $+\pi$ in azimuth, $0.65r_p$ to
$1.35r_p$ in radius, and $\pi/2-3h_p$ to $\pi/2$ in colatitude, where
$h_p=0.05$ is the disk's aspect ratio at the orbital radius of the
planet, so that $99$~\% of the disk's mass at the planet location lies
within the mesh. As we simulate only one hemisphere of the disk, we
adopt reflecting boundary conditions at the equator. At high altitude
we extrapolate the density and azimuthal velocity fields using the
analytic profiles of Appendix~\ref{sec:prof-disk-rotat} to fill the
ghost zones, whereas we use symmetric boundary conditions on the
radial velocity and antisymmetric boundary conditions on the velocity
in colatitude. Our mesh size respectively in azimuth, radius and
colatitude is $N_\phi=3290$, $N_r=367$ and $N_\theta=78$, with a
uniform spacing, so that the cells are approximately cubic at $r=r_p$,
with an edge length of $1.9\cdot 10^{-3}r_p$. The planetary potential has
the form:
\begin{equation}
  \label{eq:51}
  \Phi_p(\boldsymbol r) = -\frac{GM_p}{\sqrt{|\boldsymbol r-\boldsymbol r_p|^2+\epsilon^2}},
\end{equation}
where $\epsilon$, a softening length used to avoid a divergence at the
planet location, is set to $4\cdot 10^{-3}r_p$, which is $8$~\% of the
pressure scale height, or two cell sizes. No viscosity is used in the
calculations.  The planet to star mass ratio is $q=10^{-5}$, so the
planet has $8$~\% of the thermal mass $h_p^3M_\star$.

The disk's initial conditions are given by
$v_r\equiv v_\theta\equiv 0$, while the density and azimuthal velocity
are given respectively by Eqs.~(\ref{eq:83})
and~(\ref{eq:84}). Damping boundary conditions \citep{valborro06} are
used in the radial direction only, over $10$~\% of the radial extent
of the mesh (\emph{i.e.} over annuli of width $0.07r_p$), both at the
inner and outer edges. Each of our calculations is carried out over
$20$~orbital periods of the planet. We do not use a temporal tapering
of the planetary mass upon the planet insertion in the disk. The
manner in which the planet is turned on can have an impact on the flow
at larger time in two dimensions \citep{2015MNRAS.446.1026O}, since
some vortensity can be created by shocks and trapped in a closed
region around the planet, but no such region exists in three
dimensions \citep{2015MNRAS.447.3512O}, and our setup does not have
the resolution to capture processes that happen on a scale of the
planetary Bondi radius.

Our fiducial calculation is one for which $\alpha=3/2$, which
corresponds to a (nearly) vanishing vortensity gradient. Our
background disk is therefore identical to that of
\citet{2015arXiv150503152F}. The main differences are, in our case, a
smaller planetary mass, and a much coarser resolution at the planet
location.

In addition to this fiducial calculation, we have run ten others
calculations in which we vary the vortensity gradient, defined by
Eq.~(\ref{eq:48}). Overall we have $11$ runs with a vortensity
gradient ranging from $-1.5$ to $1.5$ by increments of $0.3$.

\subsection{Separatrix sheet and stagnation points}
\label{sec:separ-sheet-stagn}
We perform all our analyses at $t=20$~orbital periods of the planet, the date
$t=0$ corresponding to the planet insertion in the disk.

We determine the position of the separatrices by integrating the
path of fluid elements starting at a given colatitude and radius $r$, with
azimuth $\phi =\pm 1$. The integration is carried out downstream for
$\phi(r-r_c)>0$, and upstream otherwise. The trilinearly interpolated
value of each component of the velocity field is used for the
integration. For each colatitude, the starting radius of the widest
horseshoe streamline is found with a dichotomic search to machine
accuracy.

\begin{figure}
  \centering
  \includegraphics[width=\columnwidth]{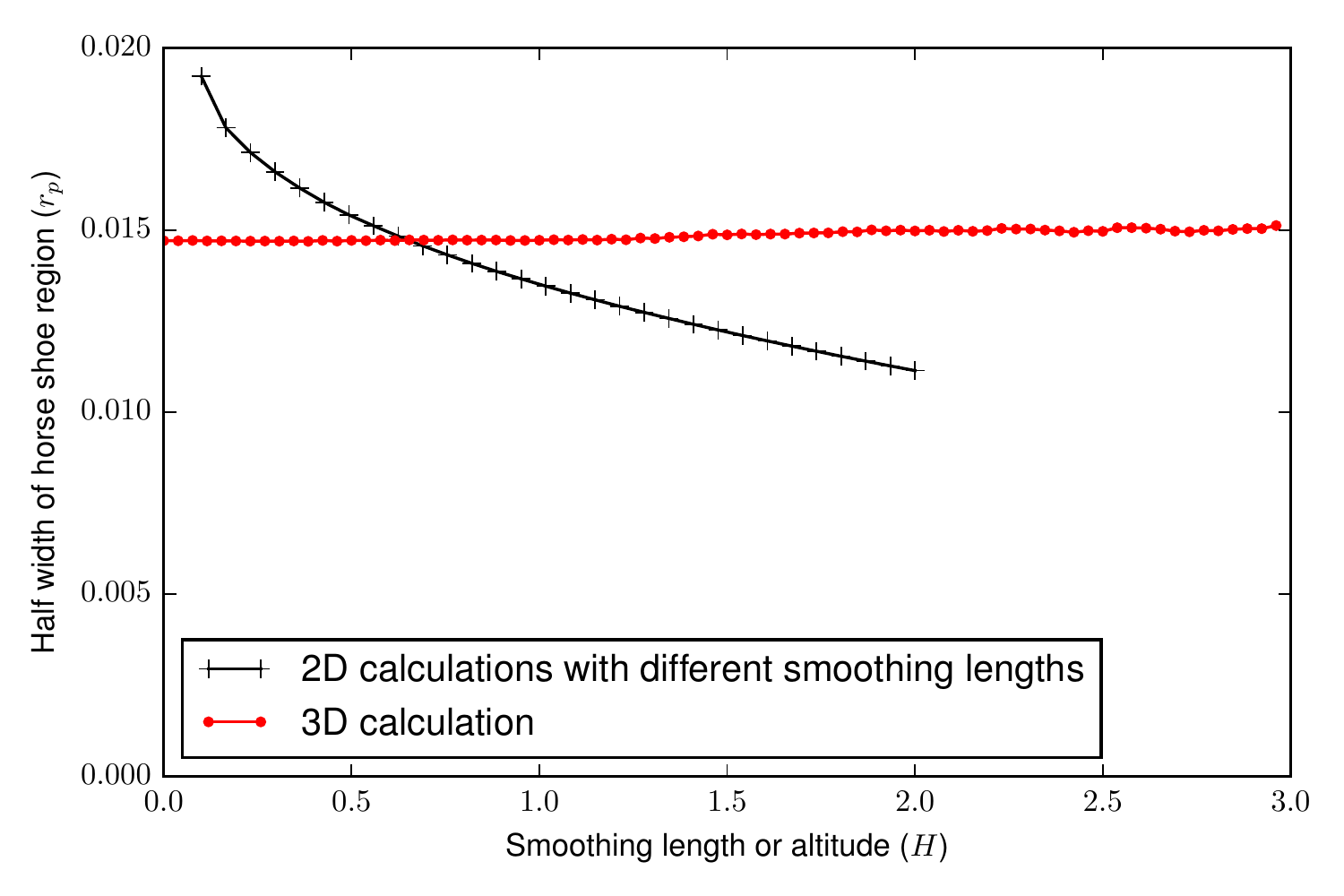}
  \caption{Width of the horseshoe region as a function of altitude
    $(\pi/2-\theta)r_p$, expressed in pressure scale heights $H$. The
    value reported is the arithmetic mean of the width measured at
    $\phi=1$ and $\phi=-1$ (using a dichotomic search of the
    separatrices position; the dots represent the altitudes at which
    this search was performed). The black crosses show the width
    measured in auxiliary two-dimensional calculations, in which case
    the $x$-axis represent the softening length used for the
    potential, in units of $H$. This figure should be compared to
    figure~3 of \citet{2015arXiv150503152F}.}
  \label{fig:2dvs3d}
\end{figure}

We represent on Fig.~\ref{fig:2dvs3d} the width of horseshoe region as
a function of the distance to the midplane. As expected from
section~\ref{sec:cons-about-topol}, the width is nearly constant, to
within a few percent, over the whole vertical extent of the
computational domain, which covers three vertical scale heights. Also
shown in this figure is the width of the horseshoe region from 2D
calculations with different potential's softening lengths. We see that
they match for a softening length of $0.65H$. Interestingly, this
value is comparable to the softening length required to match the two-
and three-dimensional Lindblad torques
\citep{masset02,2012AA...546A..99K}.  These results are to be
compared with those of \citet{2015arXiv150503152F}, who considered a
marginally sub-thermal mass planet, and who find the horseshoe width
reproduced by 2D calculations with a softening length $\sim 0.35H$.
The average value of the front and rear widths measured in our
fiducial run is:
\begin{equation}
  \label{eq:52}
  x_s=0.0147r_p 
\end{equation}

In order to determine the location of the stagnation points, we
proceed as outlined in section~\ref{sec:cons-about-topol}, i.e. by
determining the manifolds of decreasing dimension obtained by imposing
successively that $v_\phi=0$, then also $v_r=0$, and finally also
$v_\theta=0$.  Care must be taken in this procedure with the
staggering of the velocity components in the FARGO3D code. Namely, for
each of the $N_\theta$ conical slices of our mesh, we determine the
location at which $v_\phi$ cancels out (the intersection of the
corotation sheet with the cone $\theta=\theta_k$, with
$\theta_k=\pi/2-3h_p(k+1/2)/N_\theta$ and $k\in[0,N_\theta-1]$). This
location turns out to be unique, for our setup and resolution, and
results in a function of $r$, for any given azimuth
$(\phi_i)_{i\in[0,N_\phi-1]}$: $r=r_c(\phi_i,\theta_k)$. Next we
linearly interpolate $v_r$ at the locations
$[\phi_i, r_c(\phi_i,\theta_k), \theta_k]$. For each value of $k$
(within each slice in colatitude), we thus construct a sequence
$v_r^i$. We then seek the root(s) in $\phi$ of the piecewise linear
function that has value $v_r^i$ for $\phi=\phi_i$. We thus obtain, for
each slice in colatitude, the location at which the bilinearly
interpolated values of the azimuthal and radial velocities
simultaneously vanish. By varying $k$, we identify ``filaments of
horizontal stagnation'', nearly vertical curves in which the azimuthal
and radial velocities cancel out. These filaments are not streamlines,
since they can have some tilt in azimuth and radius, whereas the
velocity field on them is purely along the colatitude direction, by
construction\footnote{If they were streamlines, they would be circles
  centered on the star and contained in vertical planes, since their
  velocity would be exclusively along the colatitude direction. This
  is not possible because, by definition, they are contained in the
  corotation sheet, which has cylindrical symmetry. A stagnation
  filament depends on the coordinate system and does not have,
  contrary to the stagnation points, a physical meaning \emph{per
    se}.}. There are not sets of stagnation points either, because in
general $v_\theta$ does not vanish on these filaments, except at
specific locations which are determined in our last step, which
consists in constructing for each filament the sequence of
$v_\theta^k$, the bilinearly interpolated values of $v_\theta$ where
the filaments intersect the zone edges in colatitude. As previously,
we then seek the roots of the piecewise linear function that coincides
with $v_\theta^k$ for $\theta=\pi/2-3h_pk/N_\theta$. This gives us the
location of the stagnation points, where the three velocity components
simultaneously vanish. As a sanity check, we verify that the
trilinearly interpolated value of each velocity component is indeed
very small at the location of each stagnation point. We overall
identify $96$~stagnation points in our fiducial run at $20$~orbits,
most of them relatively far for the planet, especially in the tadpole
regions. Of special interest is the filament that lies near the
planet, almost at the intersection of the separatrix sheets. This
filament contains three stagnation points: one at the midplane, and
two others within the first pressure scale height.

\begin{figure}
  \centering 
  \includegraphics[width=\columnwidth]{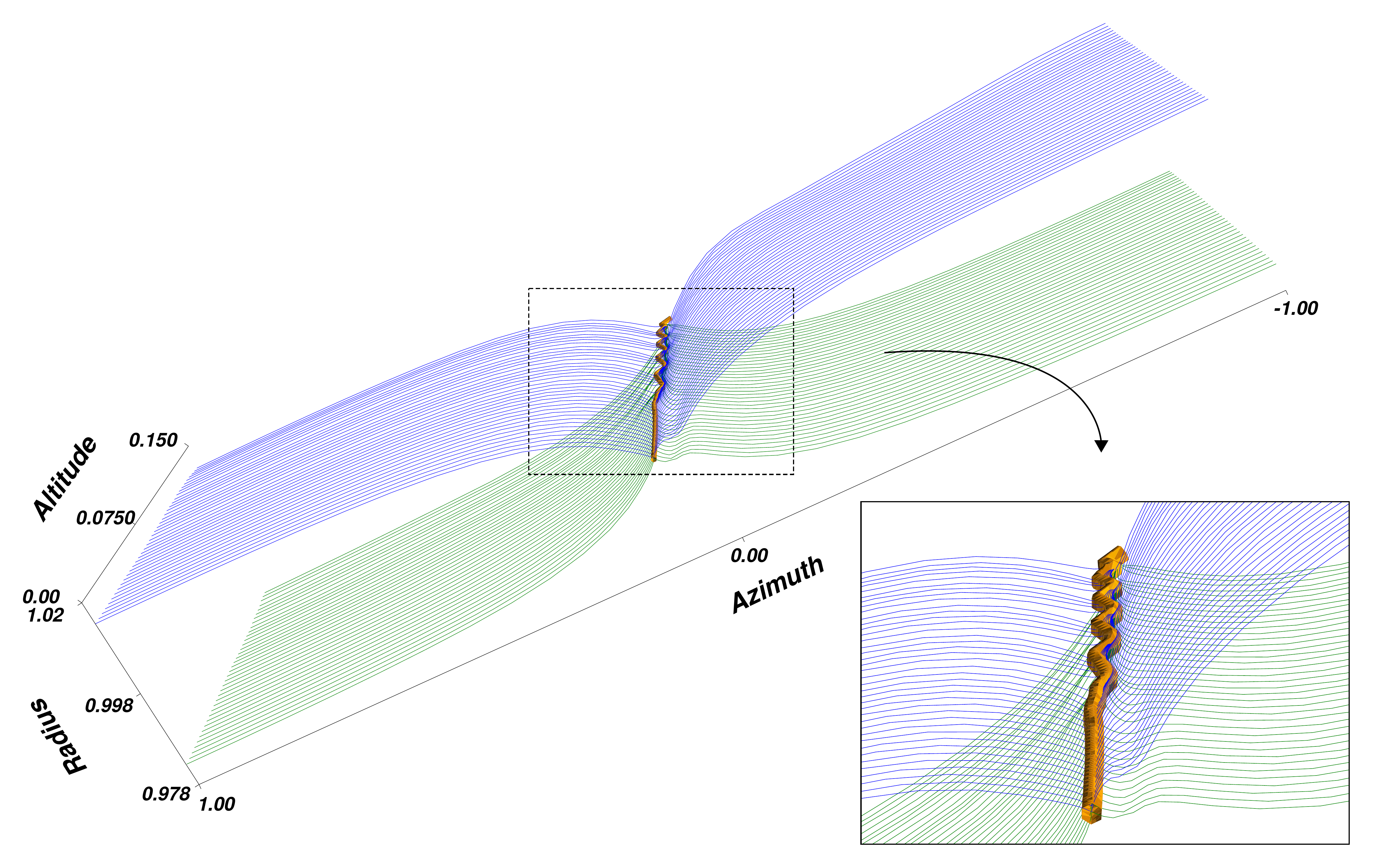}
  \caption{Three-dimensional view of the separatrices of the horseshoe 
    region (widest horseshoe streamlines). The axes represent the 
    cylindrical coordinates $(R,\phi,Z)$. For legibility purposes, the 
    radial coordinate has been stretched by a factor of 2, while the 
    azimuthal coordinate has been compressed by a factor of 5. The 
    orange tube at the streamlines intersection represents a filament 
    of horizontal stagnation, in which three stagnation points can be 
    identified (see main text for details).}
  \label{fig:v3d}
\end{figure}

We show in Fig.~\ref{fig:v3d} the separatrix sheets of the horseshoe
region. Although the streamlines of these sheets exhibit significant
vertical motion toward stagnation points in the vicinity of the
stagnation filament, as expected from the considerations of
section~\ref{sec:cons-about-topol}, they never quite reach these
points: we find that an accuracy higher than double precision would be
required to better constrain the starting radius of a separatrix
streamline that would reach the vicinity of the stagnation points. Our
assertion of section~\ref{sec:cons-about-topol} that Bernoulli's
invariant is uniform on the separatrix sheets, and equal to its value
at the corresponding stagnation point therefore needs to be examined
in more detail.

\subsection{Bernoulli's invariant in our fiducial calculation}
\label{sec:bern-invar-our}
\begin{figure}
  \centering 
  \includegraphics[width=\columnwidth]{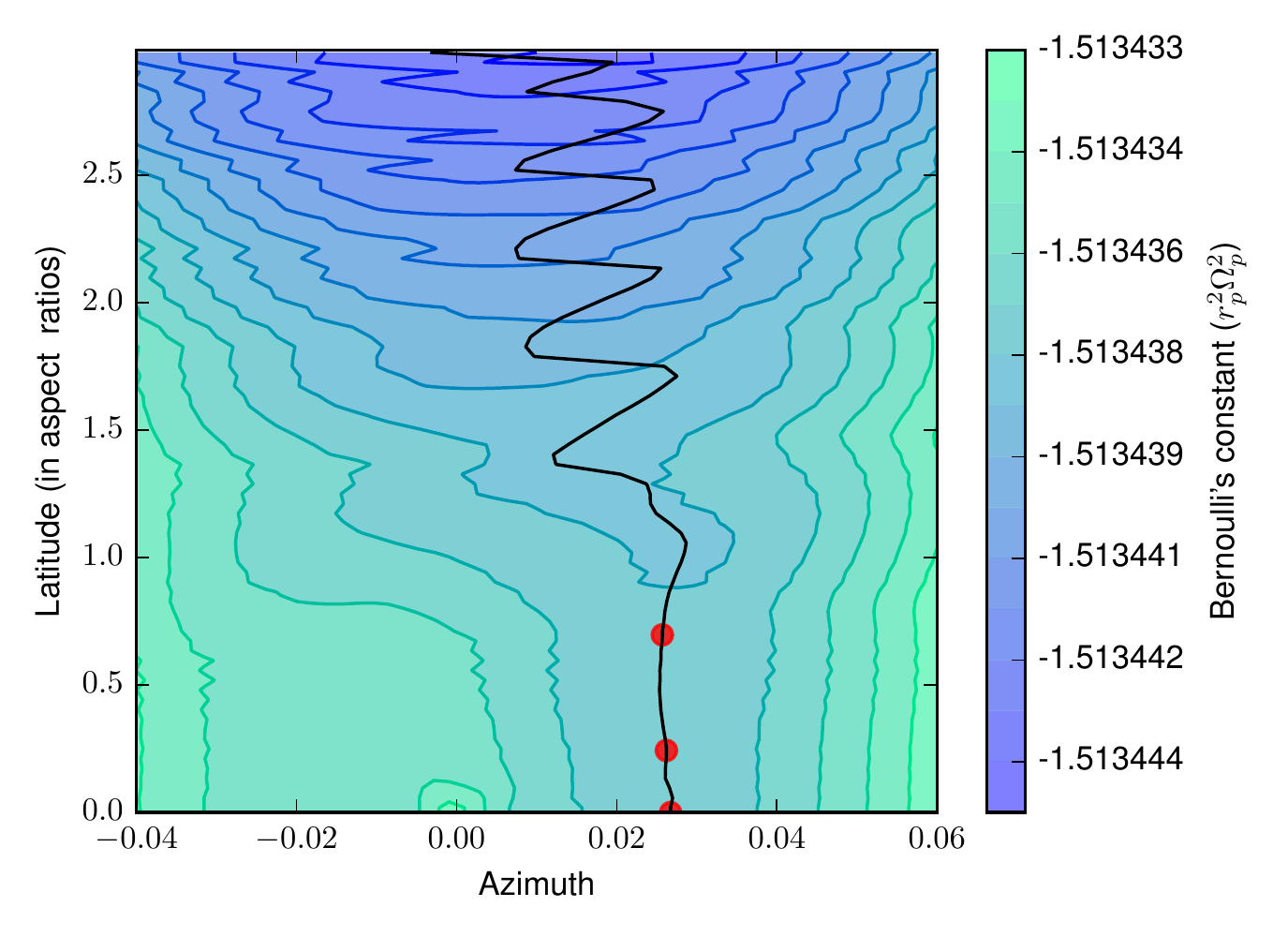}
  \caption{Bernoulli's invariant in the corotation sheet. The
    horizontal axis represents $\phi$ and the vertical axis represents
    $(\pi/2-\theta)/h$. The map shown corresponds to the linearly
    interpolated value of Bernoulli's invariant at
    $r_c(\phi,\theta)$. The black line shows the filament of
    horizontal stagnation (see text for details), and the three red
    dots on this filament are stagnation points.}
  \label{fig:berncor}
\end{figure}

\begin{figure}
  \includegraphics[width=\columnwidth]{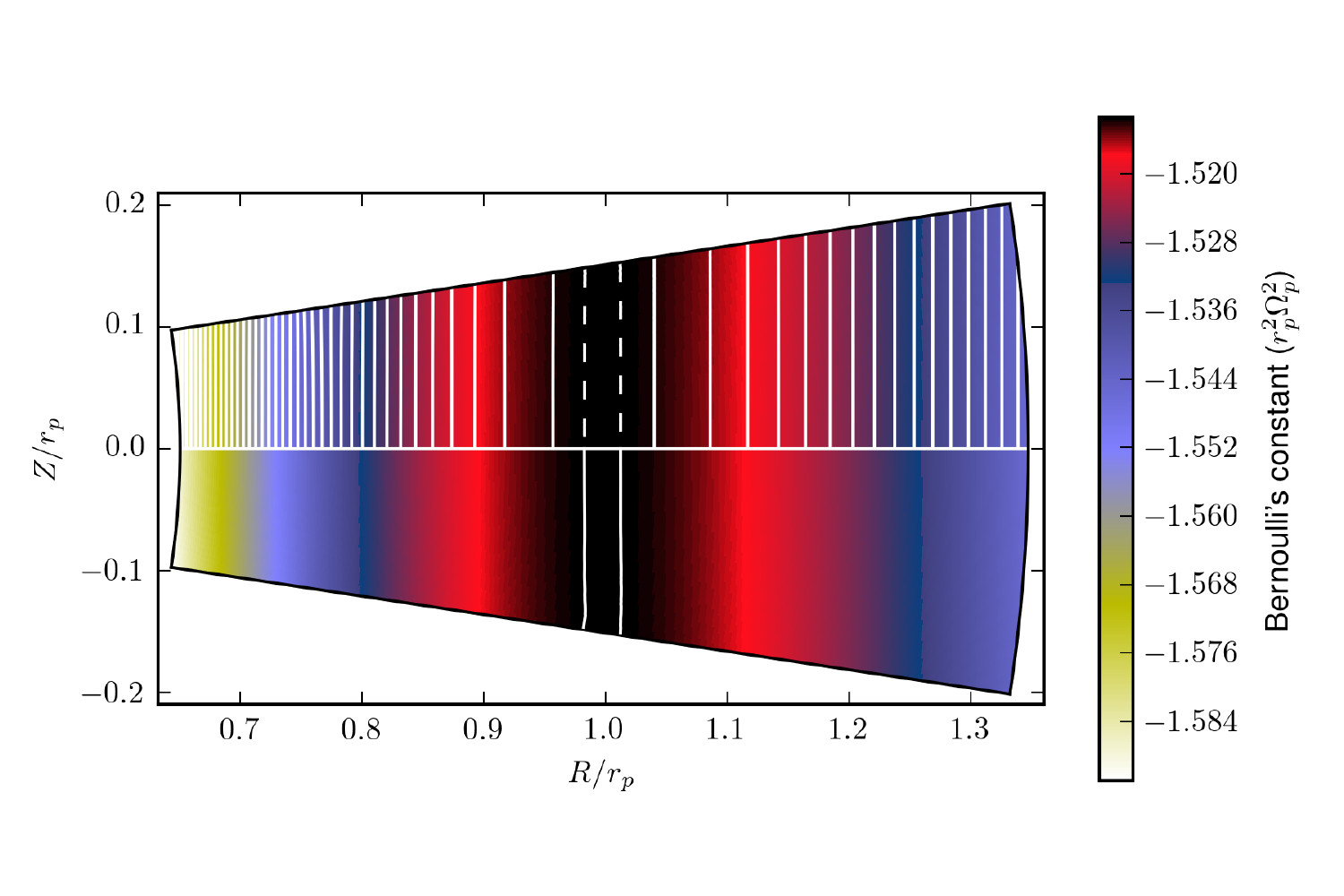}  
  \caption{\label{fig:meridbern}Meridional slice ($R,Z$) showing
    isocontours of Bernoulli's invariant in the upper half, and the
    separatrix sheet in the lower half, at azimuth $\phi=+1$~rad, for
    the whole computational domain. The dashed lines show isocontours
    of $B_s$, value measured for Bernoulli's invariant at the
    stagnation points (see text for details). As could be expected
    from Eq.~\eqref{eq:21} or Eq.~\eqref{eq:46}, Bernoulli's invariant
    is maximal at corotation.} \end{figure}

Fig.~\ref{fig:berncor} shows the value of Bernoulli's invariant
in the corotation sheet, in the planet vicinity. The horizontal
stagnation filament is shown in this figure. Bernoulli's invariant is
found to vary by less than $10^{-6}r_p^2\Omega^2$, at the
filament location, over the first pressure scale height vertically,
and it decays by $6-7\cdot 10^{-6}r_p^2\Omega_p^2$ over the next two
scale heights. This overall variation is to be put in contrast to the
drop of Bernoulli's invariant between corotation (at large distance
from the planet) and separatrices,
given by Eqs.~\eqref{eq:46} and~\eqref{eq:52}, which amounts to
$-8\cdot 10^{-5}r_p^2\Omega_p^2$.
The variation over the first scale-height is therefore at the percent
level, whereas the variation at higher altitude is at most ten percent
of the drop of Bernoulli's invariant between corotation and
separatrices.  The value of Bernoulli's invariant at the three
stagnation points on the filament are respectively, by order of
increasing distance to the midplane:
\begin{eqnarray}
  B_s^1 = -1.51343737 r_p^2\Omega_p^2\nonumber \\
  B_s^2 = -1.51343732 r_p^2\Omega_p^2\nonumber \\
  B_s^3 = -1.51343750 r_p^2\Omega_p^2\nonumber
  \end{eqnarray}
  In order to better assess to which extent Bernoulli's invariant can
  be regarded as uniform over the separatrix sheets, we define
  $B_s=-1.5134374 r_p^2\Omega_p^2$, an average of the values quoted
  above, and we compare the isosurfaces $B_J\equiv B_s$ to the
  separatrix sheets. Fig.~\ref{fig:meridbern} shows a global view of a
  meridional slice of the disk. The upper half of the figure shows
  isocontours of Bernoulli's invariant at $\phi=1$~rad. The lower half
  shows the separatrix sheets at same azimuth. The cylindrical
  symmetry of the separatrix sheet and iso-Bernoulli surfaces is
  readily apparent, and we find an excellent coincidence between the
  radii of the separatrices and the radii of the isosurfaces of
  $B_s$. Here we use a subscript $s$ to convey that $B_J$ has to be
  evaluated at a stagnation point. In
  section~\ref{sec:corot-torq-integr}, we used a similar notation to
  represent the (unique) value of $B_J$ on a separatrix sheet. Our
  numerical results confirm our expectation of
  section~\ref{sec:cons-about-topol} that both values coincide, hence
  it is legitimate to use the same notation $B_s$ for both values.

\begin{figure}
  \centering
  \includegraphics[width=\columnwidth]{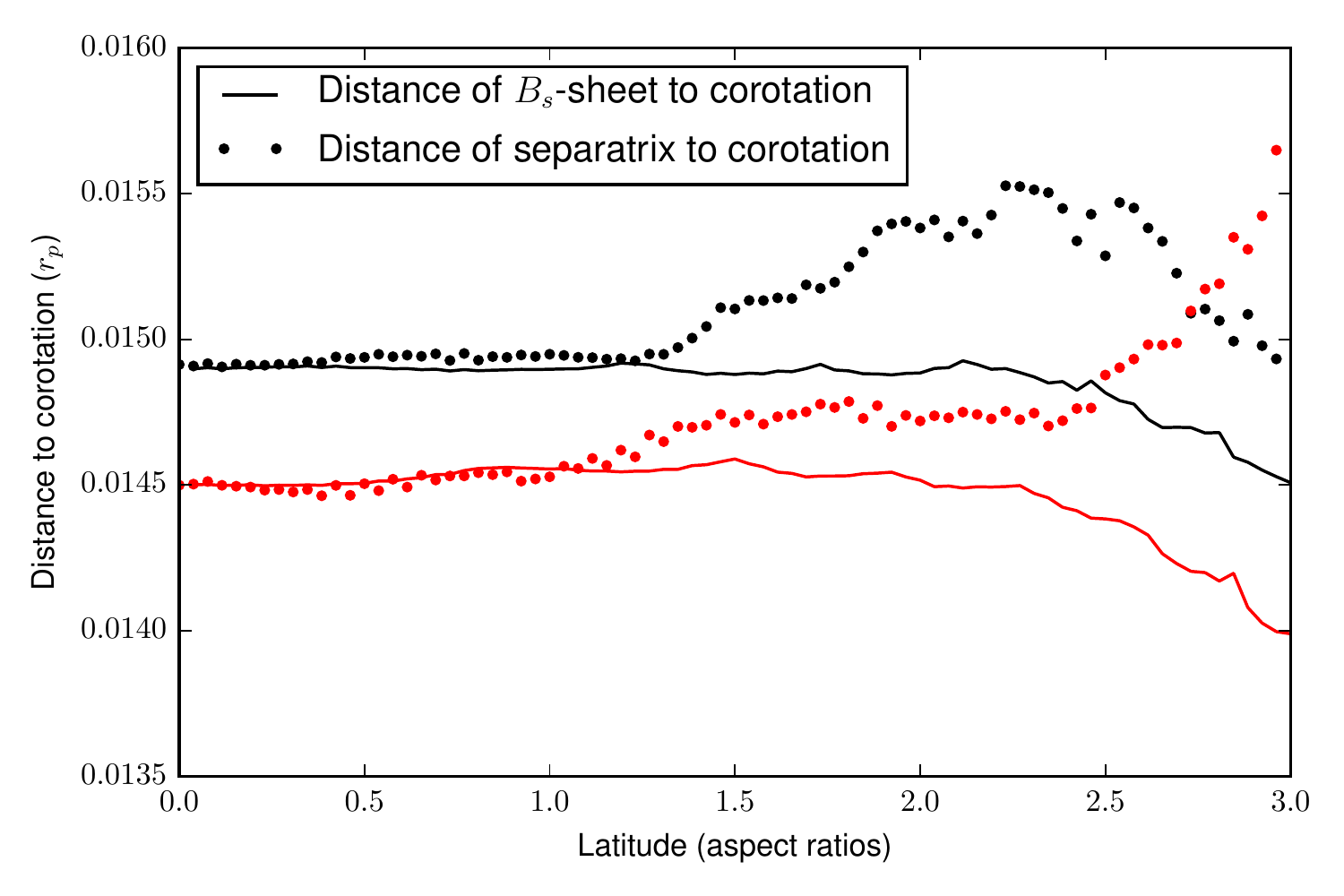}
  \caption{Detailed view of the upstream separatrix versus isosurface of
    Bernoulli's invariant, for the front part ($\phi = 1$~rad, in
    black) and the rear part ($\phi=-1$~rad, in red). The $x$-axis
    represents $(\pi/2-\theta)/h_p$, and the $y$-axis the distance to
    corotation at constant colatitude, in semi-major axis of the planet.}
  \label{fig:detailisobs}
\end{figure}

Fig.~\ref{fig:detailisobs} shows a detailed comparison of the upstream
separatrix distance to corotation with the distance of the isosurface
$B_J\equiv B_s$. A nearly perfect agreement is found over the first
pressure scale height, with residual errors far below the mesh
resolution. It is likely that the trilinearly interpolated values of
Bernoulli's invariant on the one hand, and of the velocity on the
other hand, fulfill Eq.~(\ref{eq:14}) with a high accuracy, but we
have not investigated the reasons for this nearly perfect match in
detail. Nonetheless, this remarkable agreement lends confidence in the
fact that Bernoulli's invariant can be regarded as uniform on the
separatrix sheet, and equal to the value $B_s$ that it has on the
stagnation filament. The agreement, as can be seen in
Fig.~\ref{fig:detailisobs}, is not so good at higher altitudes (albeit
with a discrepancy between Bernoulli's isosurface and the separatrices
still below the mesh resolution).  This corresponds to the altitudes
at which Bernoulli's invariant shows a small departure from the
uniform value that it has over the first pressure scale height (see
Fig.~\ref{fig:berncor}), which suggests that the flow has not reached
a fully steady state at higher altitudes $20$~orbital periods after
the insertion of the planet. Our streamline integration shows indeed
that the high altitude fluid elements near the separatrix spend a
considerable amount of time near the horizontal stagnation filament,
where they execute significant vertical excursions at very small
velocity (thereby contributing to homogenize Bernoulli's invariant
along this filament.) We will come back to the timescale of high
altitude U-turns in section~\ref{sec:width-at-midplane}, and we will
see that they can last much longer than in the midplane.  We note that
we could not have taken a significantly larger time after the planet
insertion to perform our analysis, because the first effects of phase
mixing are expected around $t=40$~orbital periods for the horseshoe
zone width that we measured.

We note in Fig.~\ref{fig:berncor} that over the first pressure
scale-height, the filament of horizontal stagnation is almost
vertical. We could have defined stagnation filaments by the
requirement $v_R=v_\phi=0$ (that is, the first two components of the
velocity in a \emph{cylindrical} coordinate system vanish).  The
filament thus defined could nearly be a streamline connecting the
stagnation points, directly evidencing why they share same value of
Bernoulli's invariant.

\subsection{Total torque and horseshoe drag}
\label{sec:total-torq-hors}
Our following step is to check whether Eq.~(\ref{eq:49}) gives a
reliable estimate of the horseshoe drag. This expression requires the
prior knowledge of $x_s$, the half-width of the horseshoe region. We
use for this purpose the set of simulations that we performed with
different vortensity gradients, and we assume that the width of the
horseshoe region, as in the two-dimensional case, does not sensitively
depend on the vortensity gradient \citep{cm09} and is given by
Eq.~\eqref{eq:52}.

\begin{figure}
  \centering
  \includegraphics[width=\columnwidth]{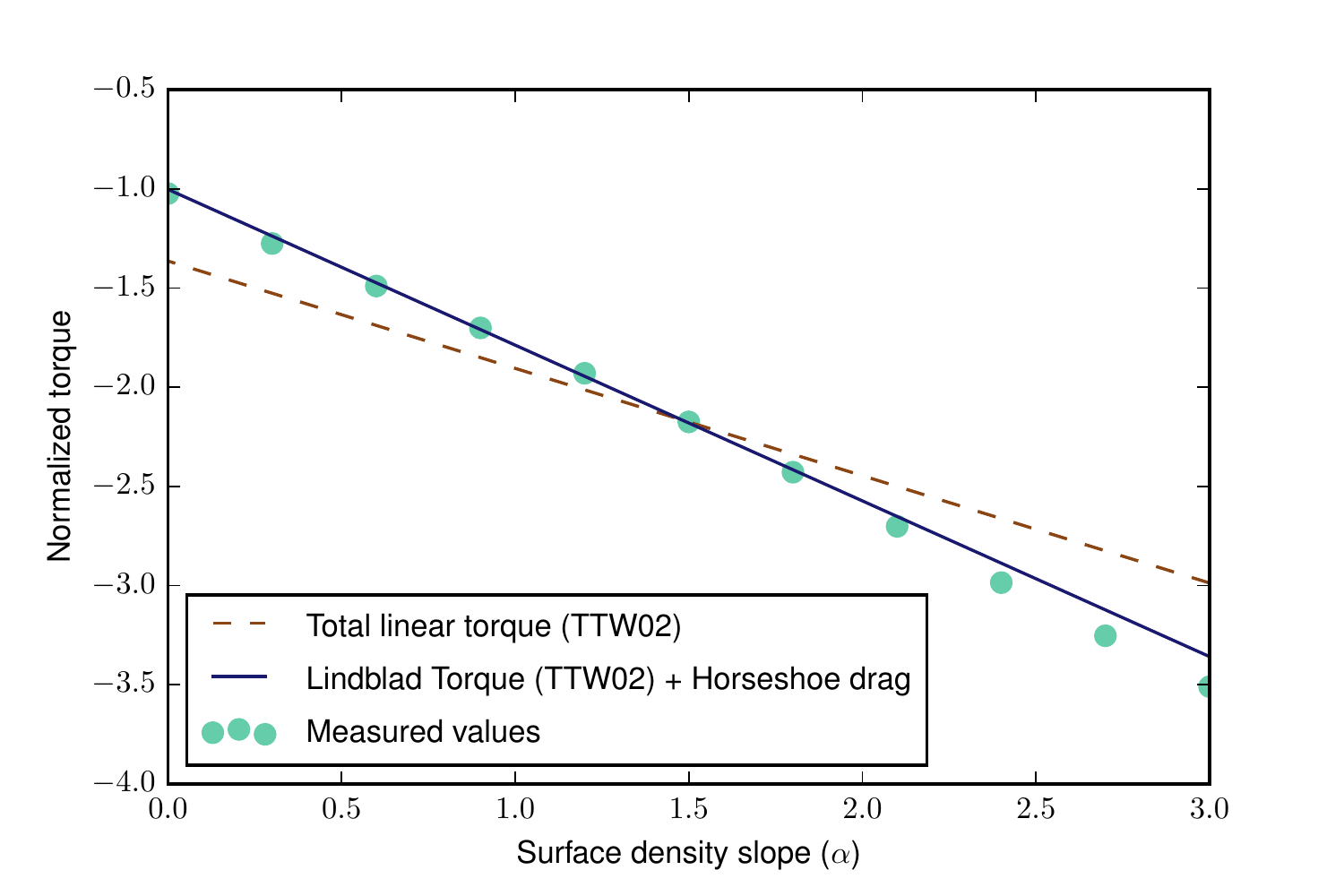}
  \caption{Total torque as a function of the surface density
    slope. The dashed line shows the total torque estimate by
    \citet{tanaka2002}, while the solid line shows the Lindblad torque
    estimate of \citet{tanaka2002}, plus the non-linear corotation
    torque estimate provided by the horseshoe drag formula of
    Eq.~(\ref{eq:49}), in which the width of the horseshoe region is
    provided by Eq.~\eqref{eq:55}.}
\label{fig:tqvsal}
\end{figure}

We show in Fig.~\ref{fig:tqvsal} the value measured for the torque in
our 11~calculations, as a function of the surface density slope, and
we compare it to two analytical expressions: one of them is the total
torque estimate obtained in the linear regime by TTW02, which reads:
\begin{equation}
  \label{eq:53}
  \Gamma_\mathrm{TTW02}=-(1.364+0.541\alpha) \Sigma_p\Omega_p^2r_p^4q^2h_p^{-2},
\end{equation}
and the other one is the sum of the linear
estimate of the Lindblad torque of TTW02, and the non-linear estimate
of the corotation torque of Eq.~(\ref{eq:49}), in which we set ${\cal
  V}=\frac 32-\alpha$.
The Lindblad torque estimate of TTW02 is: 
\begin{equation}
  \label{eq:54}
\Gamma^\mathrm{LR}_\mathrm{TTW02}=-(2.340-0.099\alpha) \Sigma_p\Omega_p^2r_p^4q^2h_p^{-2}.
\end{equation}
The value obtained in
our calculations are much more compatible with the second estimate.
This result extends to the three-dimensional case the findings of
\citet{2009arXiv0901.2265P} that the corotation torque
eventually becomes non-linear at all planetary masses, even those
which are largely sub-thermal. In addition, it shows that the
horseshoe drag formula of Eq.~(\ref{eq:49}) gives an acceptable
estimate of the non-linear corotation torque.

\subsection{Width of the horseshoe region}
\label{sec:width-hors-regi}
The half width that we have found for the horseshoe region, given by
Eq.~(\ref{eq:52}), is only $3.5$ times larger than the softening
length. One can therefore wonder whether the horseshoe region could be
substantially wider in the limit of a vanishing softening length.  In
order to answer this question, we have undertaken two additional
calculations with the same setup as our fiducial one, except for the
softening length, which was set respectively to $0.06H$ and $0.10H$,
instead of our fiducial value of $0.08H$. We find that the horseshoe
half-width displays a nearly linear relationship with the softening
length, with a very small slope, and that the width extrapolated for a
vanishing softening length is only $0.8$~\% wider than in our fiducial
calculation: $x_s=0.0148r_p$.

This width is marginally smaller than the widely used estimate
derived for two-dimensional disks with a softening length of $0.3H$
for the potential, which reads $x_s=1.16r_p(q/h)^{1/2}$ \citep{mak2006}.
Here, we find a slightly different numerical factor, about $10$~\%
smaller, for the half-width of the horseshoe region:
\begin{equation}
  \label{eq:55}
x_s=1.05r_p\sqrt{\frac{q}{h}}.
\end{equation}
These findings are consistent with those of
\citet{2015MNRAS.452.1717L}, who also observe a $10$~\% smaller width
for the horseshoe region than the standard two-dimensional estimate
for sub-thermal mass planets.

\subsection{Relationship between perturbed effective potential and
  horseshoe width}
\label{sec:relat-betw-pert}
In the limit of a small planetary mass, the location of all stagnation
points tend to the corotation of the unperturbed disk, since
$v_\phi=R(\Omega_0-\Omega_p)+v_\phi'$, where $v_\phi'$ is the perturbed velocity due to
the planet, which scales with $q$ when the planetary mass is
sufficiently small. We restrict our discussion to this limiting case,
which is tantamount to assuming that the corotation sheet is not
significantly distorted by the introduction of the planet, at the
location of the stagnation points. Denoting $\boldsymbol r_s$ the
location of a stagnation point, we can write Bernoulli's constant at
$\boldsymbol r_s$ prior to the planet insertion, which reads:
\begin{equation}
  \label{eq:56}
  B_{J,0}=\Phi_*(\boldsymbol r_s)-\frac 12
  R_c\Omega_p^2+\eta_0(\boldsymbol r_s)=B_c,
\end{equation}
and after the planet insertion, once a steady state is reached in the
planet frame:
\begin{equation}
  \label{eq:57}
  B_J=\Phi_*(\boldsymbol r_s)+\Phi_p(\boldsymbol r_s)-\frac 12
  R_c\Omega_p^2+\eta(\boldsymbol r_s)=B_s.
\end{equation}
We therefore have:
\begin{equation}
  \label{eq:58}
  B_s-B_c=\Phi_p(\boldsymbol r_s)+\eta'(\boldsymbol r_s),
\end{equation}
where $\eta'=\eta-\eta_0$ is the perturbed enthalpy. Using a notation
similar to \citet{mak2006}, we write:
\begin{equation}
  \label{eq:59}
  \tilde\Phi'=\Phi_p+\eta',
\end{equation}
 which corresponds to the perturbed value of the
effective potential of Eq.~\eqref{eq:10}. Using Eq.~\eqref{eq:46}, and
specifying to the Keplerian case, for which $A_p=-3\Omega_p/4$ and
$B_p=\Omega_p/4$, we obtain:
\begin{equation}
  \label{eq:60}
  x_s=\frac{1}{\Omega_p}\sqrt{-\frac 83\tilde\Phi'(\boldsymbol r_s)},
\end{equation}
which is the same as Eq.~(12) of \citet{mak2006}. The value of the
perturbed effective potential at the stagnation point is directly
related to the width of the horseshoe region, and may be used to infer
the latter from numerical simulations, without resorting to streamline
analysis.

\subsection{Tilt angle}
\label{sec:tilt-angle}
\begin{figure}
  \centering 
  \includegraphics[clip=true,trim=1.5cm 0 0 0,width=\columnwidth]{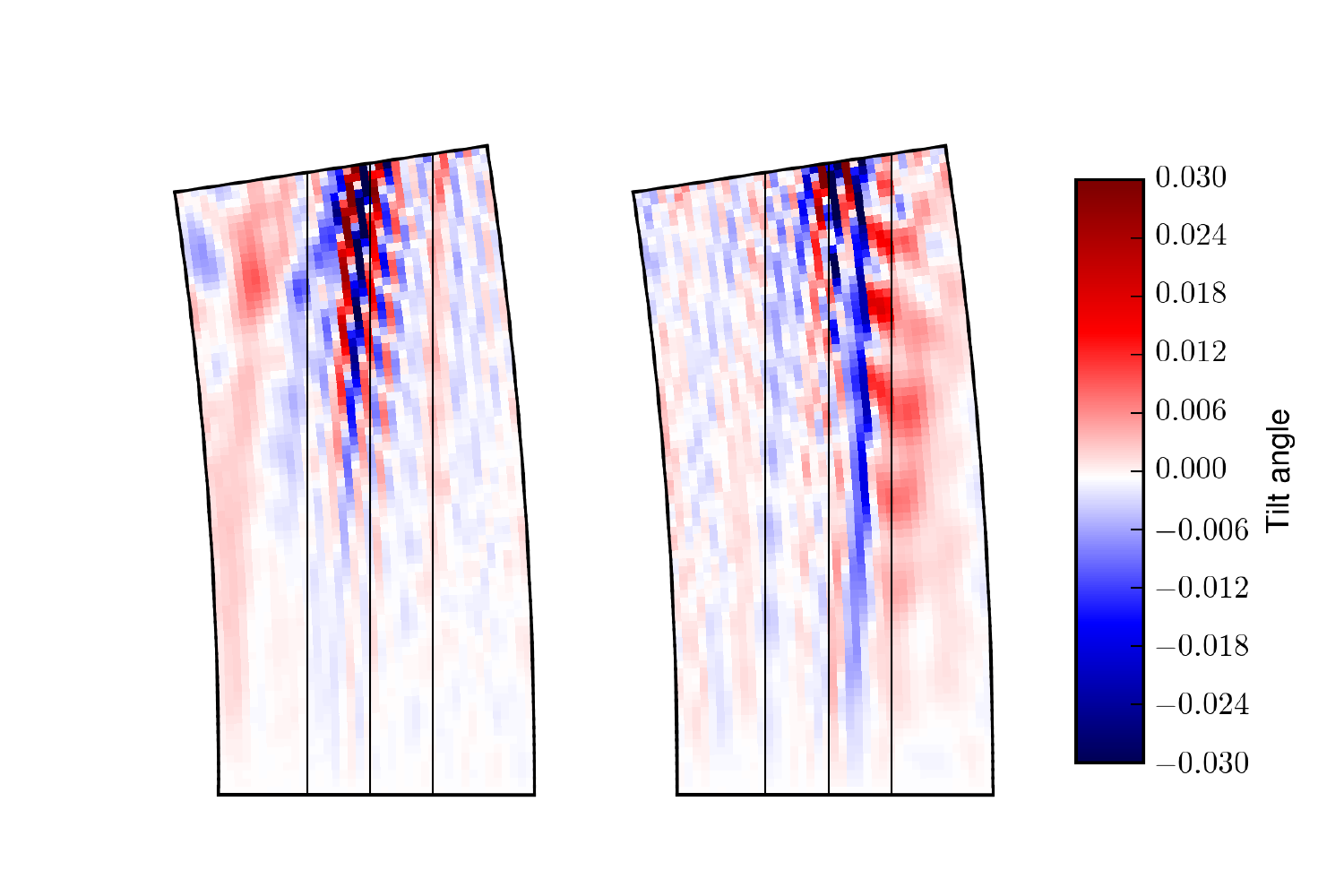}
  \caption{Tilt angle ($\zeta_R/\zeta_z$) of the vorticity vector, at
    $\phi=+1$~rad (left) and $\phi=-1$~rad (right). The three thin
    vertical lines represent, from left to right, the inner
    separatrix, the corotation and the outer separatrix.}
  \label{fig:tilt}
\end{figure}

In the derivation of Eq.~(\ref{eq:49}), we had to assume that the
vortex tubes had a small tilt angle with respect to the vertical
direction in the downstream flow. We examine here how justified this
assumption was. Fig.~\ref{fig:tilt} shows the tilt angle in the radial
direction (the tilt angle in the azimuthal direction does not feature
in our torque derivation owing to the assumption that at large
azimuthal distance the flow variables have a vanishing azimuthal
derivative.) We see on this figure that slightly larger tilt angle
values are systematically found in the downstream flow (inner side for
$\phi=1$~rad, outer side for $\phi=-1$~rad). Nevertheless, the values
observed are very small, largely below $10^{-2}$ over most of the
horseshoe flow, especially over the first pressure scale height, whose
contribution dominates the horseshoe drag. Our assumption was
therefore largely satisfied for our fiducial run.

\subsection{Potential vorticity in the midplane}
\label{sec:potent-vort-midpl}
\begin{figure*}
  \centering
  \includegraphics[width=.5\textwidth]{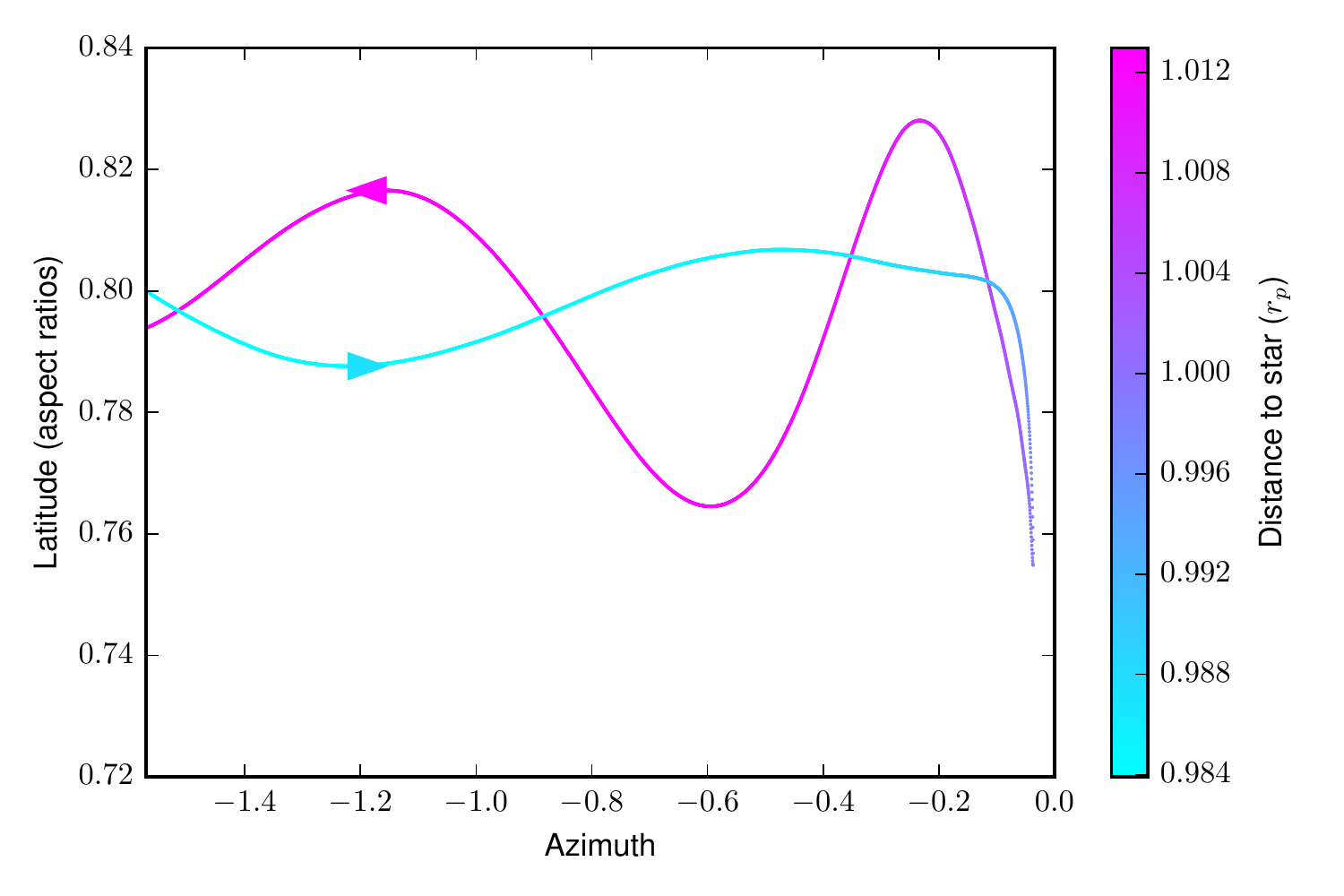}\includegraphics[width=.5\textwidth]{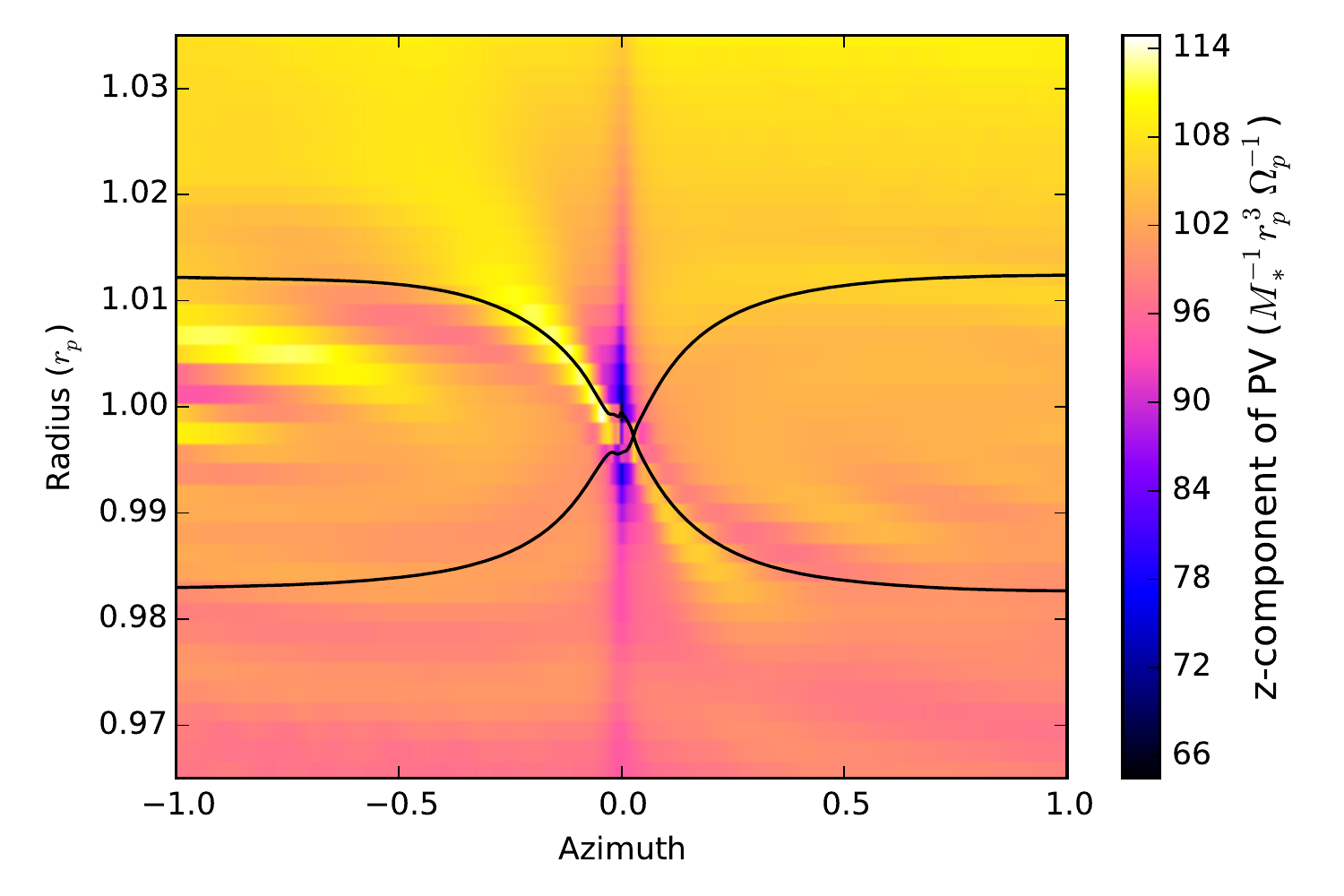}
  \caption{Horseshoe streamline near the separatrix, as seen from the
    star (left) and PV in the midplane (right). The color of the
    streamline on the left plot represents the distance to the
    star. The cyan part is the upstream one, while the purple part is
    downstream. The streamline decays toward the planet at the tip of
    the U-turn, as it tends to go toward the uppermost stagnation point of
    Fig.~\ref{fig:berncor}. It then emerges at a different altitude
    and oscillates vertically in the downstream flow. One can see
    on the right plot the imprint of these oscillations in the
    downstream flow, at the disk midplane, as the yellow and purple
    stripes. The black line shows the intersection of the separatrix
    with the midplane.}
  \label{fig:pvmid}
\end{figure*}

We finally mention that, as expected, the PV is not conserved in the
flow. This is illustrated in Fig.~\ref{fig:pvmid} where we see stripes
in the downstream distribution of the vertical component of PV (at the
midplane), which cannot be accounted for by advection of the initial
distribution by the horseshoe flow. These stripes can be traced back
to vertical oscillations of the material upon execution of their
U-turn. Compression (expansion) of material along a vortex tube
(\emph{i.e.} essentially in the vertical direction, in our case)
decreases (increases) the PV (as can be readily seen, for instance,
from Ertel's theorem by taking for $\psi$ any curvilinear abscissa
along the vortex tube).

\section{Summary}
\label{sec:summary}
We have shown that the horseshoe region of a low-mass planet in a
globally isothermal disk has a constant width with altitude and a
cylindrical shape, in line with the recent results of
\citet{2015arXiv150503152F} for nearly thermal mass planets. We
interpret this fact as the consequence of \emph{a)} the conservation
of Bernoulli's invariant, \emph{b)} the fact that this constant must
be uniform over the whole separatrix sheet (both boundaries of the
horseshoe region), and \emph{c)} the fact that Bernoulli's invariant
does not depend on the altitude, but only on the cylindrical radius,
in unperturbed globally isothermal disks. We note that in our
numerical results, the horseshoe width is constant with altitude
within a few percent, over the three scale lengths covered by our
computational domain, and within one percent over the first pressure
scale height. In contrast, the results of \citet{2015arXiv150503152F}
show a weak bulge at the $\sim 10$~\% level at the disk midplane. It
is yet to be determined whether this difference arises from the
different masses (\citet{2015arXiv150503152F} consider a nearly
thermal mass, whereas we examined here the flow around a planet with
only $8$~\% of the thermal mass), or whether it arises from the
transient horseshoe flow unveiled by \citet{2015arXiv150503152F}, that
the analysis exposed here does not resolve. In any case, the results
of \citet{2015arXiv150503152F} are indicative of some disruption of
the horseshoe separatrix near the midplane, since otherwise, as we
have shown here, the separatrix sheets should have a strictly
cylindrical shape. We further discuss this in
section~\ref{sec:relat-with-flow}.

In addition, we find that the horseshoe drag in three-dimensional,
globally isothermal disks, given by Eq.~\eqref{eq:49}, has same
expression as in two-dimensional disks, when expressed in terms of the
width of the horseshoe region. In particular, the horseshoe drag is
found to scale with the vortensity gradient. Our analysis of section
\ref{sec:torque-expression} shows that, in a three-dimensional disk,
the vortensity should be defined by Eq.~\eqref{eq:50}.

The three-dimensional case therefore essentially retains the simplicity of
the two-dimensional case. We find that the half-width of the horseshoe
region, for sub-thermal mass planets, has the form given by
Eq.~(\ref{eq:55}), which can be recast, for an arbitrary adiabatic
index $\gamma$, as:
\begin{equation}
  \label{eq:61}
  x_s=1.05\gamma^{-1/4}r_p\sqrt{\frac qh}.
\end{equation}
This result is in good agreement with the recent findings of
\citet{2015MNRAS.452.1717L}, who also find that the width of the
horseshoe region in three-dimensional situations is $\sim 10$~\%
smaller than previously envisioned from two-dimensional calculations
with the customary value $\epsilon=0.3H$ for the softening length of
the potential.  One consequence of this somewhat unexpected result is
discussed in section~\ref{sec:effect-x_s-saturated}.  Incidentally, we
find that the value of the softening length that leads to
two-dimensional horseshoe regions with the width of Eq.~\eqref{eq:61}
is $\epsilon=0.65H$, a value close to that required to match the
Lindblad torques of two- and three-dimensional disks.

\section{Discussion}
\label{sec:discussion}
\subsection{Comparison to the linear corotation torque}
\label{sec:comp-line-corot}
Using Eq.~(\ref{eq:49}) and (\ref{eq:55}), we can recast the
non-linear corotation torque expression as:
\begin{equation}
  \label{eq:62}
  \Gamma_\mathrm{CR}
=0.88{\cal V}\Sigma_p\Omega_p^2r_p^4q^2h_p^{-2}.
\end{equation} 
Assuming that to lowest order ${\cal V}=3/2-\alpha$, we have:
\begin{equation}
  \label{eq:63}
  \Gamma_\mathrm{CR}
\approx(1.32-0.88\alpha)\Sigma_p\Omega_p^2r_p^4q^2h_p^{-2}.  
\end{equation} 
This expression is to be compared with the linear corotation
torque of TTW02, which reads:
\begin{equation}
  \label{eq:64}
  \Gamma_\mathrm{CR}^\mathrm{lin}
\approx(0.976-0.64\alpha)\Sigma_p\Omega_p^2r_p^4q^2h_p^{-2},
\end{equation}
showing that the non-linear corotation torque is $\sim 1.36$ times
larger than the linear corotation torque. This statement is
reminiscent of the two-dimensional case where a similar ratio is found
between the non-linear and linear estimates of the corotation torques
\citep{2010ApJ...723.1393M,pbk11}.

The linear estimate of TTW02 vanishes for $\alpha=1.525$,
slightly above the value $3/2$ that one would naively expect. The
vortensity gradient in a globally isothermal disk is actually not
$3/2-\alpha$. Using Eq.~(\ref{eq:84}), we can show that it has the
expression: \begin{equation}
  \label{eq:65}
  {\cal V}=\frac 32\left(1+\frac 32h^2\right)-\left(1-\frac 32 h^2\right)\alpha,
\end{equation}
where we neglect terms of order $h^4$ and above. The vortensity
gradient therefore vanishes for $\alpha = 3/2+(9/2)h^2$ (which in our
fiducial disk is $\alpha=1.511$) and so does the non-linear corotation
torque, which scales \emph{exactly} with the vortensity gradient in
our analysis.

\subsection{On the width at the midplane and at higher altitude}
\label{sec:width-at-midplane}
One might have expected the width of the horseshoe region to decrease
at higher altitudes, since the gas there is subjected to a reduced
gravity from the planet, compared to the midplane. Similarly, one
might have expected the width in the midplane to be similar to the
width of the two-dimensional case in the limit of a vanishing
softening length, since in this plane the motion is
two-dimensional. This two-dimensional width, however, is much larger
than the width reported here: instead of the value given by
Eq.~(\ref{eq:61}), it is $\sim 1.7r_p\sqrt{q/h}$
\citep{2009arXiv0901.2263P,2013MNRAS.428.3526O}. Therefore two
questions arise: why is the horseshoe region so narrow at the
midplane, and why is it so wide at large altitudes~?

\subsubsection{Width at the midplane}
\label{sec:width-at-midplane-1}
\begin{figure}
  \centering
  \includegraphics[width=\columnwidth]{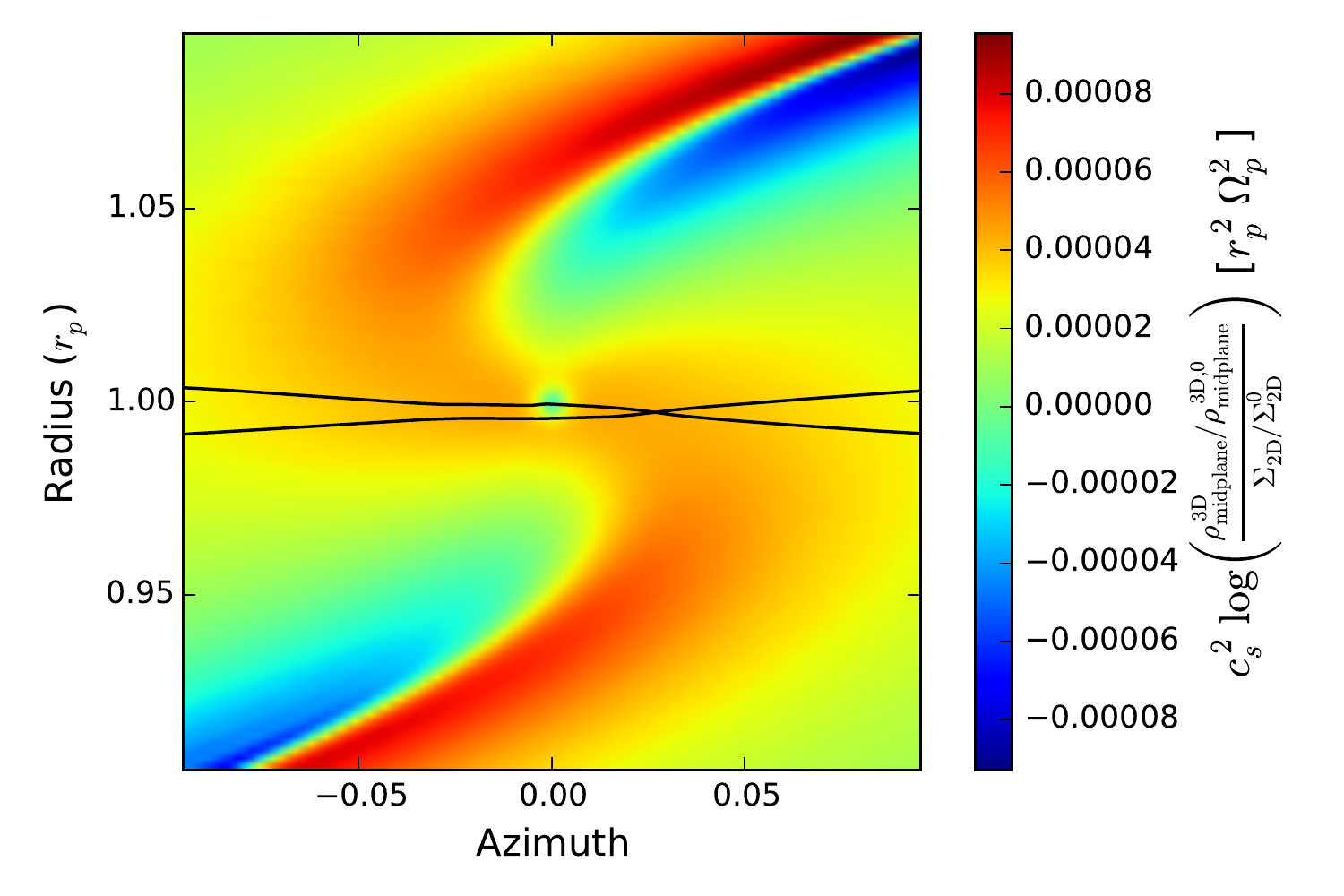}
  \caption{Difference of perturbed enthalpy between the
    three-dimensional case (at the midplane) and a two-dimensional
    case with same parameters. A positive value is found for
    $r\sim r_p$.  Incidentally, we note an enthalpy excess upstream of
    the spiral wake, followed by a deficit downstream, compatible with
    a convergence of the fluid elements toward the midplane as they
    arrive at the wake, followed by an expansion.}
  \label{fig:enth23}
\end{figure}

Eq.~\eqref{eq:60} shows that the width of the horseshoe region is
determined by the value of the perturbed effective potential, given by
Eq.~\eqref{eq:59}, at the stagnation point.  The well of perturbed
effective potential is much shallower than the gravitational potential
well of the planet, as the latter is partially filled with enthalpy
(which is why the horseshoe region is much narrower than predicted by
the restricted three body problem). In the 3D case, the density at the
midplane differs from the density in a purely two-dimensional case
with same velocity field because of the vertical motion above the
midplane, which can compress the gas or allow it to expand. The
difference of perturbed enthalpy between the three-dimensional case
and a two-dimensional case with same resolution and parameters is
shown in Fig.~\ref{fig:enth23}. This difference is also the difference
of perturbed effective potentials between the two cases, since the
planetary potential is the same in the two- and three-dimensional
cases.  In the vicinity of the orbit, this difference is positive,
resulting from the trend of the streamlines to bend toward the
midplane. The effective potential well is therefore shallower in the
3D case than in the 2D case, and the horseshoe region is consequently
narrower. The actual difference of the horseshoe width between the 2D
and 3D cases cannot be deduced from Fig.~\ref{fig:enth23}, because the
stagnation point can be at a different location in each case, but the
order of magnitude of the enthalpy excess in the 3D case is compatible
with a sizable reduction of the horseshoe width with respect to the
two-dimensional case.

\subsubsection{Width at high altitude}
\label{sec:width-at-high}
As we mentioned in section~\ref{sec:bern-invar-our}, the high altitude
fluid elements originating from the vicinity of the separatrix linger
near the stagnation filament, taking thereby a large amount of time to
perform their U-turn.  This effect is responsible for maintaining a
sizable width at higher altitudes. The time it takes to perform a
horseshoe U-turn near the midplane is only a few orbital periods of
the planet \citep{bm08}, but it has to take more time at higher
altitude for the following reason: the planetary torque is smaller
than its midplane value, but a fluid element located near the
separatrix eventually exchanges with the planet the same amount of
angular momentum as that of a fluid element at the same cylindrical
radius in the midplane, because the horseshoe region has the same
width at all altitudes. The reduction factor can be large in our
fiducial case, since the typical azimuth along the stagnation
filament, $\phi \sim 0.02$, is small compared to the latitude
($\sim 0.15$~rad) of the highest fluid elements, which indicates that
the time it takes to execute a U-turn at higher altitude is large
compared to that required near the midplane.

\subsection{Considerations on saturation}
\label{sec:cons-satur}
This paper has focused on the unsaturated horseshoe drag. A detailed
study of the saturation properties will be presented elsewhere. We
can however anticipate here the following two points:
\begin{itemize}
\item As pointed out in section~\ref{sec:width-at-high}, the widest
  horseshoe U-turns are performed more slowly at large altitude. In a
  two-dimensional situation, saturation occurs as a result of phase
  mixing because streamlines located at different distances of
  corotation have different libration times. In a three-dimensional
  situation, streamlines at same distance to corotation but at
  different altitudes also have different libration times\footnote{The
    libration time is only a factor $h^{-1}$ larger than the U-turn
    time for sub-thermal mass planets \citep{bm08}. If the U-turn time
    at higher altitude is several times larger than at the midplane,
    it may then represent a sizable fraction of the overall libration
    time.}. We can therefore anticipate that the torque oscillations
  should exhibit a pattern different from those of two-dimensional
  disks \citep[see e.g.][]{2010ApJ...723.1393M}, and that the torque
  should relax faster toward its asymptotic value as a result of
  enhanced phase mixing.
\item In the evaluation of the saturated torque value, the parameter
  $x_s$ (half-width of the horseshoe region) plays a very important
  role. The torque asymptotic (or saturated) value depends on a
  competition between libration (which tends to cancel the torque, as
  it flattens the gradients of vortensity and entropy across
  corotation) and diffusive processes (viscous and thermal diffusion,
  which tend to restore the large scale gradients and the unsaturated
  torque value). Of these two processes, the one with the shortest
  timescale will win the competition.  The ratio of the diffusive to
  the libration time scales is proportional to $x_s^3$
  \citep{masset01,2010ApJ...723.1393M,pbk11}, hence the ultimate
  torque value depends very sensitively on $x_s$. An accurate estimate
  of $x_s$ such as the one we provide here is therefore crucially
  important in correctly evaluating the degree of saturation of the
  corotation torque. \citet{2010ApJ...723.1393M} suggested that the
  width of the horseshoe region was expected to depend on the
  altitude, and that a value at low altitude, where most of the mass
  resides, could be used to determine the torque value. In view of the
  results reported here, this suggestion is clearly not adequate, and
  it leads, not surprisingly, to corotation torque values too
  saturated and therefore largely sub-estimated
  \citep{2011AA...536A..77B}. Regardless of the improvements required
  for three-dimensional torque formulae, we show in
  appendix~\ref{sec:effect-x_s-saturated} that using for $x_s$ the
  value given by Eq.~(\ref{eq:61}) improves considerably the torque
  estimate of \citet{2010ApJ...723.1393M} and essentially reconciles
  it with other estimates \citep{pbk11}.
 \end{itemize}

\subsection{Relationship with the flow at sub-Bondi scale}
\label{sec:relat-with-flow}
Recent work has highlighted the characteristics of the flow at the
scale of Bondi's radius $r_B=GM_p/c_s^2$ and below
\citep{2015MNRAS.447.3512O}, and the potential impact of this flow on
the horseshoe drag \citep{2015arXiv150503152F}.
\citet{2015MNRAS.447.3512O} show that the flow can enter the Bondi
sphere at high altitude and exit near the midplane (in shear
dominated configurations) or enter the Bondi sphere near the midplane
and exit at larger altitude (in headwind dominated configurations). In
the former case, \citet{2015arXiv150503152F} have found, for a
slightly sub-thermal mass planet, that the material expelled near the
midplane can reach and cross the horseshoe's boundary. They call this
flow the transient horseshoe flow, as it involves fluid elements that
participate in a unique close encounter with the planet. This flow
destroys the horseshoe boundary (near the midplane),
invalidating our assumption of the existence of a critical surface:
streamlines of different origins merge near the edge of the downstream
horseshoe flow. Our work completely disregards these potentially very
important effects, owing to its moderate resolution. As noted by
\citet{2015arXiv150503152F}, the huge resolution needed to
resolve the flow within the Bondi sphere for planetary masses as low
as those considered here is beyond computational tractability on
present day platforms, at least for single mesh calculations, even
with non-uniform resolution. The problem of the impact of the flow at
sub-Bondi scales on the horseshoe drag should be tackled by means of
nested mesh calculations, which is a forthcoming feature of the code
that we used in this paper. \citet{2015arXiv150503152F} find that the
total torque in their calculation is still negative, but reduced in
magnitude with respect to the expected value of the linear Lindblad
torque. They perform a horseshoe drag analysis of their outcome by
summing the mass flow rate, weighted by the angular momentum jump
(which they determine by a streamline integration), over the whole set
of horseshoe streamlines. They find that the contribution of the
transient horseshoe flow to the torque is negligible, and that there
exists a net positive corotation torque due to the standard horseshoe
flow, in spite of the disk being barotropic with a vanishing
vortensity gradient. It is noteworthy that they find a non-vanishing
contribution at all altitudes (see their Fig. 15), even above the
first pressure scale height, where the flow most resembles the
situation that we described here, with a columnar structure and well
defined separatrix sheets. There are ingredients neglected in our
analysis that could play an important role, such as the radial tilt of
vortex tubes upon U-turns, or the non-conservation of Bernoulli's
constant if shocks are present in the planet vicinity. The impact of
the flow at sub-Bondi scale on the planetary torque clearly requires
further work, in which such ingredients should probably be
incorporated.

Extensions of the present work could also consider the impact of the
radial temperature gradient (in so-called locally isothermal disks) or
the role of the entropy gradient in flows with an energy equation (in
so-called adiabatic disks). The entropy gradient is known to have a
major impact on the corotation torque in three-dimensional disks
\citep{pm08, 2015MNRAS.452.1717L}, but so far its action on the
coorbital flow and the generation mechanism of the entropy-related
torque have been studied only in two-dimensional disks
\citep{mc09,pbck10}.

\acknowledgments We thank Bertram Bitsch for kindly providing us with
the simulation data described in
appendix~\ref{sec:effect-x_s-saturated}. We thank Jeffrey Fung, Pawel
Artymowicz and Yanqin Wu for insightful discussions on the transient
horseshoe drag. We also thank Gloria Koenigsberger, Jeffrey Fung,
Elena Lega and Bertram Bitsch for constructive comments on a first
draft of this manuscript, and an anonymous referee for constructive
comments which improved the quality of this manuscript.  F. Masset
acknowledges support from CONACyT grant~178377. Pablo Ben\'\i
tez-Llambay acknowledges financial support from CONICET.  We thank
Ulises Amaya Olvera, Reyes Garc\'\i a Carre\'on and J\'er\^ome
Verleyen for their assistance in setting up the GPU cluster on which
the calculations presented here have been run.

\appendix

\section{Profiles of disks in rotational and hydrostatic equilibrium}
\label{sec:prof-disk-rotat}
The standard procedure that consists in adopting a Gaussian vertical
profile of the disk density is only approximate, and has a poor accuracy at
altitudes higher than the disk's pressure scale height. Here we derive exact
relations for the disk's density and azimuthal velocity profiles, under
conditions slightly more general than those considered in
sections~\ref{sec:unsat-hors-drag} and~\ref{sec: Num}. We assume that the sound
speed is a power law of the spherical radius:
\begin{equation}
  \label{eq:66}
  c_s^2(r)= (c_s^0)^2\left(\frac{r}{r_0}\right)^{-\beta}.
\end{equation}
where $r_0$ is an arbitrary radius at which the sound speed is $c_s^0$. Such
disks are often qualified of locally isothermal. The aspect ratio has the radial
dependence:
\begin{equation}
  \label{eq:67}
  h(r) =\frac{c_s(r)}{v_K(r)}\propto r^{(1-\beta)/2},
\end{equation}
where $v_K(r)=\sqrt{GM_\star/r}$ is the circular Keplerian velocity at distance $r$
from the central mass.
We call flaring index the exponent $f$ of the power law given by Eq.~(\ref{eq:67}):
\begin{equation}
  \label{eq:68}
  f=\frac{1-\beta}{2}.
\end{equation}
For the globally isothermal disks considered in the main part of this
paper, we have $\beta=0$ and $f=1/2$.  The equations that determine
the rotational and vertical equilibria of the disk are respectively,
in spherical coordinates\footnote{In this appendix only, $v_\phi$
  denotes the azimuthal velocity in a non-rotating frame.}:
\begin{equation}
 \label{eq:69}
 -\frac{\partial_r(\rho_0 c_s^2)}{\rho_0}+\frac{v_\phi^2}{r}-\frac{GM_\star}{r^2}=0
\end{equation}
and
\begin{equation}
 \label{eq:70}
 -\frac1r\frac{\partial_\theta(\rho_0 c_s^2)}{\rho_0}+\frac{v_\phi^2}{r}\cot\theta=0
\end{equation}
If we denote $L=\log(\rho_0/\rho_{00})$, $m=v_\phi^2/c_s^2$, $u=-\log(\sin\theta)$, $v=\log(r/r_0)$ and $K=GM_\star/[c_s(r_0)^2r_0]$, we can transform Eq.~(\ref{eq:70}) into:
\begin{equation}
 \label{eq:71}
 \partial_uL+m=0
\end{equation}
and Eq.~(\ref{eq:69}) into
\begin{equation}
 \label{eq:72}
 -\beta+\partial_vL-m+K\exp(-2fv)=0,
\end{equation}
where we have made use of the assumption that the sound speed only depends on the spherical radius. Deriving Eq.~(\ref{eq:71}) with respect to $v$ and Eq.~(\ref{eq:72}) with respect to $u$, we are led to
\begin{equation}
 \label{eq:73}
 \partial_vm+\partial_um=0,
\end{equation}
from which we infer:
\begin{equation}
 \label{eq:74}
 \partial_{u^k}^km=(-1)^k\partial_{v^k}^km.
\end{equation}
The rotational equilibrium in the midplane reads, from Eq.~(\ref{eq:69}):
\begin{equation}
 \label{eq:75}
 m(u=0,v) = -\beta-\xi+K\exp(-2fv),
\end{equation}
hence, for any $k\ge 1$, we have in the midplane ($u=0$):
\begin{equation}
 \label{eq:76}
 \partial_{v^k}^km=(-2f)^kK\exp(-2fv),
\end{equation}
so that, by virtue of Eq.~(\ref{eq:74}), we have, also in the midplane:
\begin{equation}
 \label{eq:77}
   \partial_{u^k}^km=(2f)^kK\exp(-2fv),
\end{equation}
from which we can reconstruct the value of $m$ at an arbitrary height above the midplane:
\begin{equation}
 \label{eq:78}
 m(u,v)=[\exp(2fu)-1]K\exp(-2fv)+m(u=0,v)=-\beta-\xi+K\exp[2f(u-v)],
\end{equation}
which specifies the field of rotational velocity. The density field is found by integrating Eq.~(\ref{eq:71}), which yields:
\begin{equation}
 \label{eq:79}
 L=L_\mathrm{eq}+(\beta+\xi)u-K\frac{e^{-2fv}}{2f}[\exp(2fu)-1],
\end{equation}
where the~eq subscript denotes the midplane value.
Using the more conventional notation, Eqs.~(\ref{eq:78}) and~(\ref{eq:79}) read respectively:
\begin{equation}
 \label{eq:80}
 v_\phi(r,\theta)=v_K(r)\left[(\sin\theta)^{-2f}-(\beta+\xi)h^2\right]^{1/2},
\end{equation}
where $v_K(r)$ is the circular Keplerian velocity at distance $r$ from the central mass,
and
\begin{equation}
 \label{eq:81}
 \rho_0(r,\theta)=\rho_\mathrm{eq}(r) (\sin\theta)^{-\beta-\xi}\exp[h^{-2}(1-\sin^{-2f}\theta)/2f].
\end{equation}
For a ``flat'' disk, in which the temperature is inversely proportional to the
radius ($\beta=1$ and $f=0$), the integration of Eq.~(\ref{eq:71}) eventually
yields:
\begin{equation}
 \label{eq:82}
 \rho_0(r,\theta)=\rho_\mathrm{eq}(\sin\theta)^{-\beta-\xi+h^{-2}},
\end{equation}
For globally isothermal disks, Eqs~(\ref{eq:81}) and~(\ref{eq:80}) can be recast
respectively as:
\begin{equation}
  \label{eq:83}
  \rho(r,\theta)=\rho_\mathrm{eq}\sin^{-\xi}\theta\exp\left[h^{-2}\left(1-\frac{1}{\sin\theta}\right)\right]
\end{equation}
\begin{equation}
  \label{eq:84}
  v_\phi^2(r,\theta)=\frac{GM_\star}{r\sin\theta}-\xi c_s^2=\frac{GM_\star}{R}-\xi c_s^2.
\end{equation}
The rotational velocity is therefore independent of the altitude at a given
cylindrical radius in globally isothermal disks.
Finally, for $z/R \ll 1$, we have $u\approx \frac 12(z/R)^2$, 
hence Eq.~(\ref{eq:79}) can be recast in the following approximate form, when $fu\ll 1$:
\begin{equation}
 \label{eq:85}
 L\approx L_{\rm eq}-\frac 12h(r_0)^{-2}\left(\frac{r}{r_0}\right)^{-2f}\left(\frac zR\right)^2,
\end{equation}
where use has been made of the relationship $h^{-2} \gg |\xi+\beta|$. As a consequence,
we recover the well known approximation:
\begin{equation}
 \label{eq:86}
 \rho_0(z)\approx \rho_{\rm eq}\exp(-z^2/2H^2),
\end{equation}
from which we can infer the relationships
\begin{equation}
 \label{eq:87}
 \Sigma_0(r) = \sqrt{2\pi}\rho_{\rm eq}H
\end{equation}
and
\begin{equation}
 \label{eq:88}
 \alpha = \xi-1-f
\end{equation}

\section{Uniqueness of Bernoulli's invariant on a given separatrix
  sheet}
\label{sec:uniq-bern-invar}
We use \emph{reductio ad absurdum} to show that, in steady state, a
separatrix sheet cannot be connected to two stagnation points with
different values of Bernoulli's invariant. We assume a horseshoe
separatrix sheet to be connected to two stagnation points with values
of Bernoulli's invariant $B_1$ and $B_2\ne B_1$. The value of
Bernoulli's invariant on the separatrix sheet is piecewise constant:
it is equal to $B_1$ on the streamlines connected to the first point,
and it is equal to $B_2$ on the streamlines connected to the second
point. It is therefore discontinuous at the critical streamline
separating these two domains. Eq.~\eqref{eq:15} shows that the only
term that can be discontinuous in the expression of Bernoulli's
invariant is the kinetic energy (since the gravitational potential is
continuous in space, and so is the enthalpy in a barotropic fluid,
away from shocks.) At large distance from the planet, where the radial
and vertical velocities are negligible, the azimuthal velocity must be
discontinuous across the critical streamline. The radial balance,
given by Eq.~\eqref{eq:6}, reads:
\begin{equation}
  \label{eq:89}
  \frac{v_\phi^2}{R}=-\partial_R\eta-\partial_R \Phi.
\end{equation}
The potential gradient being continuous, the discontinuity of $v_\phi$
must be borne by $\partial_R\eta$. Denoting $z_c$ the altitude of a
point~$C$ on the critical streamline, we have
$\partial_R\eta|_{z_c^+} \ne \partial_R\eta|_{z_c^-}$. This precludes
the continuity of $\eta$ on any neighborhood of~$C$, which is
impossible. Our initial assumption is therefore impossible and the
separatrix sheet cannot be connected to points with different values
of Bernoulli's invariant. Numerical experiments confirm this
expectation: discontinuities in the azimuthal velocity along the
vertical direction would correspond to singular sheets of radial
vorticity, which are not observed.

\section{Effect of $x_s$ on the saturated torque value}
\label{sec:effect-x_s-saturated}
\begin{figure}
  \centering
  \includegraphics[width=.7\columnwidth]{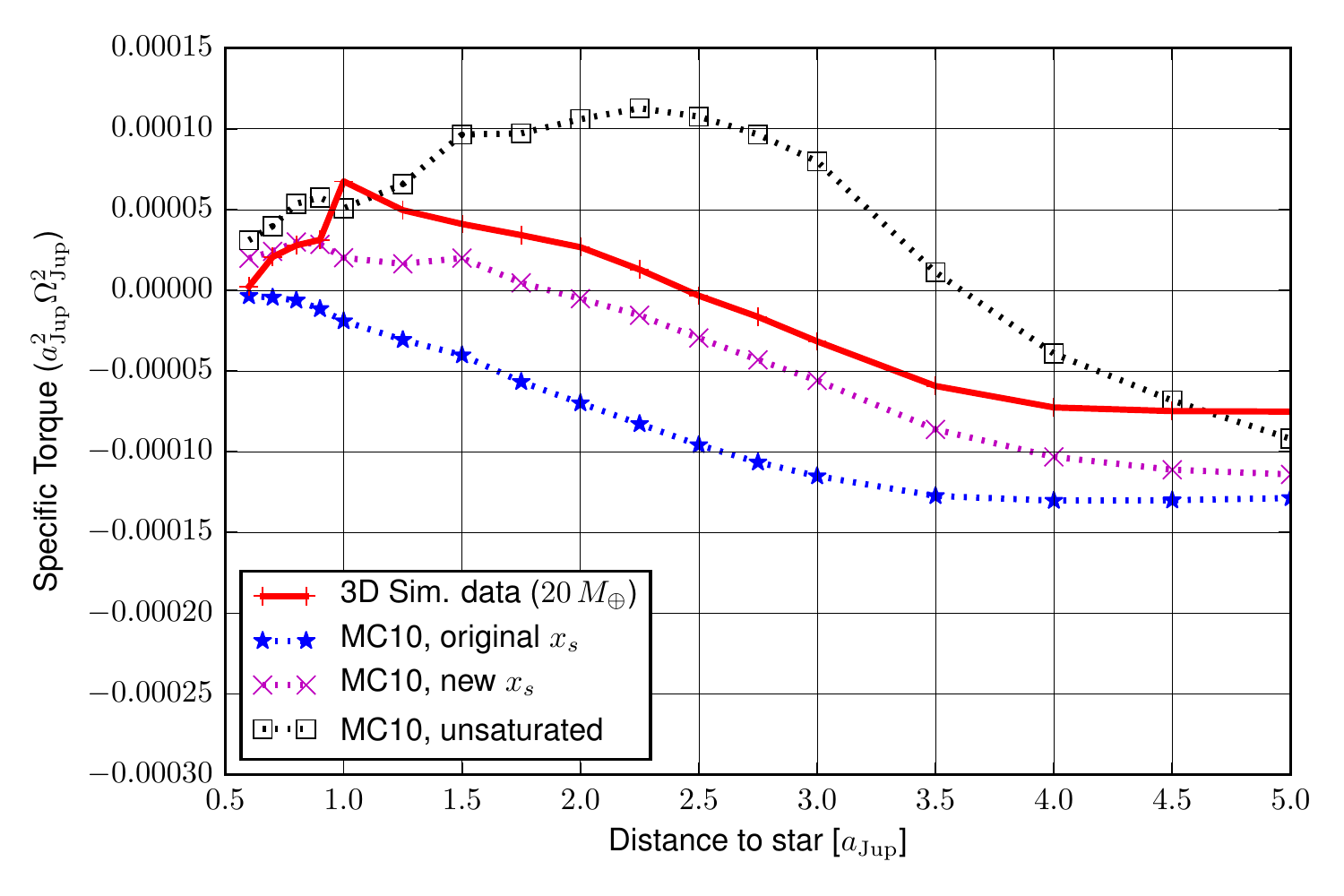}
  \caption{\label{fig:bk11}Total torque estimate with the value of $x_s$ found in this
    work (magenta curve), compared to old estimate for an over-wide
    horseshoe region (blue). This figure should be compared to Fig.~3
    of BK11. The simulation data (red curve) have been
    kindly provided by B. Bitsch.}
\end{figure}

We reproduce here the comparison performed by \citet[][hereafter
BK11]{2011AA...536A..77B} between simulations of a $20\,M_\oplus$
planet embedded in a three-dimensional radiative disk, and the torque
formula of \citet[][hereafter MC10]{2010ApJ...723.1393M}. The unique
amendment to the original formula is that we use for $x_s$ the value
provided by Eq.~(\ref{eq:61}), instead of the larger value suggested
by MC10 in their Eq.~(157). Fig.~\ref{fig:bk11} shows that the
resulting new estimate is much closer than the original one to the
simulation data. A perfect agreement should not be expected for the
large planetary mass (typically thermal) considered here, since we use
a formula valid only for largely sub-thermal mass planets.  The width $x_s'$
of the horseshoe region of a $20\,M_\oplus$ planet is likely
larger than the estimate given by Eq.~(\ref{eq:61}). The horseshoe
drag scales with $x_s^4$, and the saturation degree, for a partially
saturated torque as is the case here, scales with $x_s^{-3}$. The
torque value should therefore be underestimated by a factor $x_s'/x_s$
(where $x_s$ is given by Eq.~\ref{eq:61}), consistent with our
estimate being systematically under the simulation data. The near
compensation of these two effects (horseshoe width boost for thermal
mass planets, largely compensated for by an increased degree of
saturation) has recently been discussed by
\citet{2015MNRAS.452.1717L}, who note that torque formulae for
sub-thermal mass planets give reasonable torque estimates for
planetary masses largely beyond their domain of validity.


\begin{thebibliography}{30}
\expandafter\ifx\csname natexlab\endcsname\relax\def\natexlab#1{#1}\fi

\bibitem[{{Baruteau} \& {Masset}(2008)}]{bm08}
{Baruteau}, C., \& {Masset}, F. 2008, \apj, 672, 1054

\bibitem[{{Bitsch} \& {Kley}(2011)}]{2011AA...536A..77B}
{Bitsch}, B., \& {Kley}, W. 2011, \aap, 536, A77

\bibitem[{{Casoli} \& {Masset}(2009)}]{cm09}
{Casoli}, J., \& {Masset}, F.~S. 2009, \apj, 703, 845

\bibitem[{{D'Angelo} {et~al.}(2003){D'Angelo}, {Kley}, \& {Henning}}]{gda2003}
{D'Angelo}, G., {Kley}, W., \& {Henning}, T. 2003, \apj, 586, 540

\bibitem[{{de Val-Borro} {et~al.}(2006){de Val-Borro}, {Edgar}, {Artymowicz},
  {Ciecielag}, {Cresswell}, {D'Angelo}, {Delgado-Donate}, {Dirksen}, {Fromang},
  {Gawryszczak}, {Klahr}, {Kley}, {Lyra}, {Masset}, {Mellema}, {Nelson},
  {Paardekooper}, {Peplinski}, {Pierens}, {Plewa}, {Rice}, {Sch{\"a}fer}, \&
  {Speith}}]{valborro06}
{de Val-Borro}, M., {Edgar}, R.~G., {Artymowicz}, P., {Ciecielag}, P.,
  {Cresswell}, P., {D'Angelo}, G., {Delgado-Donate}, E.~J., {Dirksen}, G.,
  {Fromang}, S., {Gawryszczak}, A., {Klahr}, H., {Kley}, W., {Lyra}, W.,
  {Masset}, F., {Mellema}, G., {Nelson}, R.~P., {Paardekooper}, S.-J.,
  {Peplinski}, A., {Pierens}, A., {Plewa}, T., {Rice}, K., {Sch{\"a}fer}, C.,
  \& {Speith}, R. 2006, \mnras, 695

\bibitem[{{Ertel}(1942)}]{ertel42}
{Ertel}, H. 1942, Meteorol. Z., 59, 277

\bibitem[{{Fung} {et~al.}(2015){Fung}, {Artymowicz}, \&
  {Wu}}]{2015arXiv150503152F}
{Fung}, J., {Artymowicz}, P., \& {Wu}, Y. 2015, \apj, 811, 101

\bibitem[{{Goodman} \& {Rafikov}(2001)}]{gr2001}
{Goodman}, J., \& {Rafikov}, R.~R. 2001, \apj, 552, 793

\bibitem[{{Guilet} {et~al.}(2013){Guilet}, {Baruteau}, \&
  {Papaloizou}}]{2013MNRAS.430.1764G}
{Guilet}, J., {Baruteau}, C., \& {Papaloizou}, J.~C.~B. 2013, \mnras, 430, 1764

\bibitem[{{Kley} {et~al.}(2012){Kley}, {M{\"u}ller}, {Kolb}, {Benitez-Llambay},
  \& {Masset}}]{2012AA...546A..99K}
{Kley}, W., {M{\"u}ller}, T.~W.~A., {Kolb}, S.~M., {Benitez-Llambay}, P., \&
  {Masset}, F. 2012, \aap, 546, A99

\bibitem[{{Lega} {et~al.}(2015){Lega}, {Morbidelli}, {Bitsch}, {Crida}, \&
  {Szul{\'a}gyi}}]{2015MNRAS.452.1717L}
{Lega}, E., {Morbidelli}, A., {Bitsch}, B., {Crida}, A., \& {Szul{\'a}gyi}, J.
  2015, \mnras, 452, 1717

\bibitem[{{Masset}(2001)}]{masset01}
{Masset}, F.~S. 2001, \apj, 558, 453

\bibitem[{{Masset}(2002)}]{masset02}
---. 2002, \aap, 387, 605

\bibitem[{{Masset} \& {Casoli}(2009)}]{mc09}
{Masset}, F.~S., \& {Casoli}, J. 2009, \apj, 703, 857

\bibitem[{{Masset} \& {Casoli}(2010)}]{2010ApJ...723.1393M}
---. 2010, \apj, 723, 1393

\bibitem[{{Masset} {et~al.}(2006){Masset}, {D'Angelo}, \& {Kley}}]{mak2006}
{Masset}, F.~S., {D'Angelo}, G., \& {Kley}, W. 2006, \apj, 652, 730

\bibitem[{{Ogilvie} \& {Lubow}(2002)}]{og2002}
{Ogilvie}, G.~I., \& {Lubow}, S.~H. 2002, \mnras, 330, 950

\bibitem[{{Ormel}(2013)}]{2013MNRAS.428.3526O}
{Ormel}, C.~W. 2013, \mnras, 428, 3526

\bibitem[{{Ormel} {et~al.}(2015{\natexlab{a}}){Ormel}, {Kuiper}, \&
  {Shi}}]{2015MNRAS.446.1026O}
{Ormel}, C.~W., {Kuiper}, R., \& {Shi}, J.-M. 2015{\natexlab{a}}, \mnras, 446,
  1026

\bibitem[{{Ormel} {et~al.}(2015{\natexlab{b}}){Ormel}, {Shi}, \&
  {Kuiper}}]{2015MNRAS.447.3512O}
{Ormel}, C.~W., {Shi}, J.-M., \& {Kuiper}, R. 2015{\natexlab{b}}, \mnras, 447,
  3512

\bibitem[{{Paardekooper} {et~al.}(2010){Paardekooper}, {Baruteau}, {Crida}, \&
  {Kley}}]{pbck10}
{Paardekooper}, S., {Baruteau}, C., {Crida}, A., \& {Kley}, W. 2010, \mnras,
  401, 1950

\bibitem[{{Paardekooper} {et~al.}(2011){Paardekooper}, {Baruteau}, \&
  {Kley}}]{pbk11}
{Paardekooper}, S., {Baruteau}, C., \& {Kley}, W. 2011, \mnras, 410, 293

\bibitem[{{Paardekooper} \& {Mellema}(2008)}]{pm08}
{Paardekooper}, S.-J., \& {Mellema}, G. 2008, \aap, 478, 245

\bibitem[{{Paardekooper} \&
  {Papaloizou}(2009{\natexlab{a}})}]{2009arXiv0901.2265P}
{Paardekooper}, S.-J., \& {Papaloizou}, J.~C.~B. 2009{\natexlab{a}}, \mnras,
  394, 2283

\bibitem[{{Paardekooper} \&
  {Papaloizou}(2009{\natexlab{b}})}]{2009arXiv0901.2263P}
---. 2009{\natexlab{b}}, \mnras, 394, 2297

\bibitem[{{Rafikov}(2002)}]{rafikov02}
{Rafikov}, R.~R. 2002, \apj, 572, 566

\bibitem[{{Schubert} {et~al.}(2004){Schubert}, {Ruprecht}, {Hertenstein},
  {Nieto Ferreira}, {Taft}, {Rozoff}, {Ciesielski}, \&
  {Hung-Chi}}]{erttrans2004}
{Schubert}, W., {Ruprecht}, E., {Hertenstein}, R., {Nieto Ferreira}, R.,
  {Taft}, R., {Rozoff}, C., {Ciesielski}, P., \& {Hung-Chi}, K. 2004,
  Meteorologische Zeitschrift, 13, 527

\bibitem[{{Tanaka} {et~al.}(2002){Tanaka}, {Takeuchi}, \& {Ward}}]{tanaka2002}
{Tanaka}, H., {Takeuchi}, T., \& {Ward}, W.~R. 2002, \apj, 565, 1257

\bibitem[{{Ward}(1986)}]{ww86}
{Ward}, W.~R. 1986, Icarus, 67, 164

\bibitem[{{Ward}(1991)}]{wlpi91}
{Ward}, W.~R. 1991, in Lunar and Planetary Institute Conference Abstracts, 1463

\end{thebibliography}
\end{document}